\documentclass[twocolumn, journal]{IEEEtran}

\usepackage{color}
\usepackage{graphicx}
\usepackage{amsmath}
\usepackage{amssymb}

 \newcommand{\be}{\begin{equation}}
 \newcommand{\ee}{\end{equation}}
  \newcommand{\bea}{\begin{eqnarray}}
 \newcommand{\eea}{\end{eqnarray}}
 \newcommand{\ba}{\begin{array}}
\newcommand{\ea}{\end{array}}
\newcommand{\nid}{\noindent}
\newcommand{\non}{\nonumber}

\newcommand{\al}{\alpha}
\newcommand{\sg}{\sigma}

\newcommand{\bt}{\beta}

\title{Random-Training-Assisted Pilot Spoofing Detection and Secure Transmission}

\author{Xiaowen Tian,~\IEEEmembership{Student Member,~IEEE}, Ming~Li,~\IEEEmembership{Member,~IEEE}, and Qian~Liu,~\IEEEmembership{Member,~IEEE}
\thanks{Xiaowen Tian and Ming Li are with the School of Information and Communication Engineering, Dalian University of Technology, Dalian, Liaoning 116024, China, (e-mail: tianxw@mail.dlut.edu.cn, mli@dlut.edu.cn).}
\thanks{Qian Liu is with the Chair of Media Technology, Technical University of Munich, Munich 80333, Germany (e-mail: qian.liu@tum.de).}
\thanks{This paper is supported in part by the Natural Science Foundation of Liaoning Province (Grant No. 2015020043) and in part by the Fundamental Research Funds for the Central Universities (Grant No. DUT 15RC(3)121).}
\vspace{-0.0 cm}}

\begin{document}
\pagestyle{plain}

 \maketitle

\vspace{-0.0 cm}

\begin{abstract}
The pilot spoofing attack is considered as an active
eavesdropping activity launched by an adversary during the reverse channel training phase. By transmitting the same pilot signal as the legitimate user, the pilot spoofing attack is able to degrade the quality of legitimate transmission and, more severely, facilitate eavesdropping.
In an effort to detect the pilot spoofing attack and minimize its damages, in this paper we propose a novel random-training-assisted (RTA) pilot spoofing detection algorithm. In particular, we develop a new training mechanism by adding a random training phase after the conventional pilot training phase. By examining the difference of the estimated legitimate channels during these two phases, the pilot spoofing attack can be detected accurately. If no spoofing attack is detected, we present a computationally efficient channel estimation enhancement algorithm to further improve the channel  estimation accuracy. If the existence of the pilot spoofing attack is identified, a zero-forcing (ZF)-based secure transmission scheme is proposed to protect the confidential information from the active eavesdropper. Extensive simulation results demonstrate that the proposed RTA scheme can achieve efficient pilot spoofing detection, accurate channel estimation, and  secure transmission.
\end{abstract}


\vspace{-0.0 cm}
\section{Introduction}

Due to the broadcast nature of wireless medium, wireless channel is ubiquitously accessible to both legitimate and illegitimate users, and therefore vulnerable to security attacks \cite{Yulong Zou}. An eavesdropper within the coverage area of the legitimate transmitter can intercept the transmitted secure information while staying undetected. In order to maintain confidential transmission, conventional cryptographic (encryption) techniques were adopted in wireless communications aiming at disrupting the readability of the information. In recent years, physical layer security is emerging as an alternative approach to prevent eavesdropping by exploiting the physical characteristics of wireless channels.

Pioneer efforts on physical layer security were made by Shannon \cite{Shannon1}, Wyner \cite{Wyner}, Leung-Yan-Cheong and Hellman \cite{Gaussian} from information-theoretic perspective.
Based on these works, intensive studies were devoted to the development of various physical layer security techniques against eavesdropping, such as artificial noise aided security techniques \cite{AN1}-\cite{AN5}, diversity based security approaches \cite{Diversity1}, \cite{Diversity2}, physical-layer secret key generation methods \cite{keygen1}, \cite{keygen4}, and security-oriented beamforming techniques \cite{beamform1}-\cite{overview}. While these physical layer security research works are focusing on protecting the confidential information against passive eavesdropping, \textit{pilot spoofing attack} launched by an active eavesdropper arises as a more severe threat since this attack can degrade the quality of the legitimate transmission and, more vitally, facilitate eavesdropping.

In \cite{first}, Zhou \textit{et al.} investigated the pilot spoofing attack  at the first time  and analyzed its severe consequences. The pilot spoofing attack targets at the reverse channel training procedure in the time-division duplex (TDD) multi-input single-output (MISO) system, where a pilot-assisted channel estimation approach is widely adopted.
Since the pilot sequences are publicly known in general with a given protocol or standard, an active eavesdropper can transmit the same pilot sequence as the legitimate receiver in the reverse channel training phase. As a result, the channel estimation at the legitimate transmitter is biased and contains the component of illegitimate channel. If the legitimate transmitter designed the beamformer based on this biased channel estimation, the transmitted signals would turn towards the direction of the eavesdropper, which leads to performance degradation  of legitimate transmission and, more severely, information leakage to the eavesdropper.


Being aware of the serious consequences of the pilot spoofing attack, it is of significant importance and urgency for the legitimate users to detect this type of attack and take actions to protect the confidentiality of wireless communications in a practical manner.
In \cite{randomtraining}, Kapetanovi\'{c} \textit{et al.} proposed to use modified  phase-shift keying (PSK) symbols as pilots in two training slots for channel estimation.
The pilot spoofing attack can be detected by examining the phase
difference of randomly chosen PSK signals in two time slots.
However, in addition to the needs of re-designing the pilot signals
and the channel estimation process, the detection performance of this method is not very satisfactory. In \cite{ERB}, Xiong \textit{et al.} proposed an energy ratio detector which explores the power difference between the channel estimations obtained at the transmitter by uplink training and at  the receiver by downlink training. However, the training procedure with both uplink and downlink makes this approach more complex in practical implementation. Considering this problem, a novel approach was proposed in \cite{selfcontamination} based solely on uplink training. The main idea of this approach is to self-contaminate the pilot and use the information-theoretic minimum description length (MDL) algorithm to detect the spoofing attack. Unfortunately, the fore-mentioned schemes can only detect the existence of the spoofing attack but is not capable of eliminating the serious consequences in the data transmission phase. To this end, in \cite{TWTD} a two-way training detector was proposed to not only detect the spoofing attack but also avoid the information leakage by a secure transmission scheme. However, this two-way approach still requires additional downlink training which makes the protocol more complicated. Moreover, the number of pilot signal in the downlink training session is proportional to the number of the transmit antennas, which will reduce the efficiency of the system.

Inspired by these studies, in this paper we propose a novel pilot spoofing detector based solely on uplink training with the assist of random training signals. Particularly, we develop a new training mechanism by adding a random training phase after the conventional pilot training
phase. In the channel estimation procedure, the legitimate user
transmits two sets of training signals sequentially, the pre-designed pilot signal during the pilot phase and then the random
signal during the random phase. By examining the difference of estimated legitimate channels during these two phases, a novel random-training-assisted (RTA) detection algorithm is proposed to identify the existence of the pilot spoofing attack. If no spoofing attack is detected, we present a computationally efficient channel estimation enhancement algorithm to further improve the channel estimation accuracy. If the existence of the pilot spoofing attack is identified, the proposed RTA spoofing detection algorithm  can obtain the estimates of both legitimate and illegitimate channels. Based on these knowledge, we also introduce a zero-forcing (ZF)-based secure transmission method to protect the confidential information against
the active eavesdropper by maximizing the signal-to-noise ratio (SNR) at the legitimate receiver while forcing the SNR at the eavesdropper to zero.

The main contributions of our work are summarized in the following five aspects:
\begin{itemize}
\item  We present a novel random-training-assisted (RTA) spoofing detection algorithm which is based solely on uplink training and does not require significant changes on neither the design of the pilot signals nor the procedure of the channel estimation.

\item We derive the closed-form expression of the test statistic's
probability density function (PDF) and provide theoretical performance analysis of the proposed spoofing detector.

\item For no spoofing attack case, a simple channel estimation enhancement algorithm is presented in order to improve the accuracy of the
channel estimation.

\item For spoofing attack case, we also introduce a ZF-based secure transmission method to protect the confidential information from the active eavesdropper.

\item Extensive simulation results demonstrate that our proposed RTA
algorithm can achieve efficient pilot spoofing detection, accurate
channel estimation, as well as secure data transmission.
\end{itemize}

The rest of this paper is organized as follows.
In Section \ref{sec: model}, we introduce the signal model and illustrate the serious consequences caused by the pilot spoofing attack.
In Section \ref{sec: ourscheme}, we develop our RTA spoofing detector elaborately and then provide the theoretic performance analysis of it.
After introducing a channel estimation enhancement
algorithm in Section \ref{sec: enhencement}, a ZF-based secure transmission method is presented in section \ref{sec: secrecy transmission}. Simulation studies in section \ref{sec: sim} illustrate the efficiency of our proposed RTA spoofing detector. Finally, some conclusions are drawn in section \ref{sec: conclusion}.

The following notation is used throughout this paper. Boldface
lower-case letters indicate column vectors and boldface upper-case
letters indicate matrices; $\mathbb{C}$ denotes the set of all
complex numbers; $(\cdot)^T$ and $(\cdot)^H$  denote  the transpose  and
transpose-conjugate operation, respectively;  $\mathbf{I}_L$ is the
$L\times L$ identity matrix; $\mathfrak{Re} \{\cdot\}$ denotes the
real part of a complex number; $\mathrm{sgn}\{ \cdot\}$ denotes
zero-threshold quantization; $\mathbb{E} \{ \cdot \}$ represents
statistical expectation. Finally, $| \cdot |$ and $\| \cdot \|$ are the scalar magnitude and vector norm, respectively.

\vspace{-0.0 cm}
\section{System Model and Problem Formulation}
\label{sec: model}

\vspace{-0.0 cm}
\begin{center}
\begin{figure}[t]
\begin{center}
\includegraphics[width= 2.5 in]{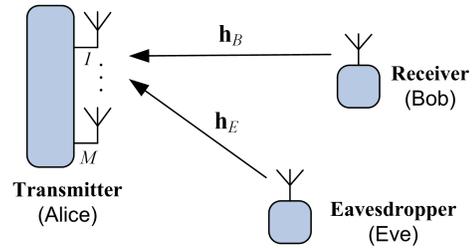} \vspace{-0.3 cm}
\caption{The MISO wiretap channel model with uplink training scheme.} \label{fg:sysmodel}\vspace{-0.0 cm}
\end{center}
\end{figure}
\end{center}
\vspace{-0.0 cm}

We consider a typical MISO wiretap system, in which the legitimate transmitter Alice is equipped with $M$ $(M\geq2)$ antennas and both the legitimate receiver Bob and the eavesdropper Eve are equipped with single antenna, as shown in Fig. \ref{fg:sysmodel}. The legitimate Bob-to-Alice channel $\mathbf{h}_B \in \mathbb{C}^{M \times 1}$ and the illegitimate Eve-to-Alice channel $\mathbf{h}_E \in \mathbb{C}^{M \times 1}$ are modeled as
\begin{equation}
\mathbf{h}_B = \sqrt{\al_B}\mathbf{\overline{h}}_B,
\end{equation}
\begin{equation}
\mathbf{h}_E = \sqrt{\al_E}\mathbf{\overline{h}}_E,
\end{equation}
respectively.
Let $\al_B$ and $\al_E$ represent the large scale (long-term) fading components (e.g. shadowing and path-loss) for the legitimate and illegitimate channels, respectively.
$\mathbf{\overline{h}}_B, \mathbf{\overline{h}}_E \in \mathbb{C}^{M \times 1}$ denote the small scale (short-term) fading coefficients (e.g. multi-path effect), and each element of $\mathbf{\overline{h}}_B, \mathbf{\overline{h}}_E$ is circularly symmetric complex Gaussian (CSCG) distributed with zero-mean and unit-variance. Under the condition of a TDD system, $\mathbf{h}_B$ and $\mathbf{h}_E$ are both reciprocal for the uplink and downlink channels. Moreover, $\mathbf{h}_B$ and $\mathbf{h}_E$ are independent from each other.

For the consideration of better performance in the legitimate transmission, Alice adopts beamforming approach to transmit signals in the direction that yields the best quality. In order to design the beamformer, it is essential for Alice to acquire the correct knowledge of $\mathbf{h}_B$ during the reverse (uplink) training phase, in which Bob sends the pre-designed pilot signals to Alice. This makes the legitimate transmission under dangerous threat due to the pilot spoofing attack in which Eve sends the same pilot signals.

In the following, we will first investigate the case in which Eve does not launch the pilot spoofing attack and build the basic signal model of channel estimation and beamforming at Alice.
Then we will describe the scenario where Eve launches the pilot spoofing attack, illustrate the serious consequences of it and attach urgent importance to the detection of such attack.

\vspace{0.5 cm}

\subsection{Channel Estimation and Beamforming without Spoofing}

For the purpose of acquiring the CSI, Bob sends the pilot signals $b(n),n=1,\ldots,N$, with power $P_B$, $N$ is the total number of pilot signals. The received signal at Alice in the $n$th time slot is
\begin{equation}
\mathbf{y}_A(n) = \sqrt{P_B} \mathbf{h}_B b(n) + \mathbf{z}(n), n=1,\ldots,N,
\label{eq:3}
\end{equation}
where $\mathbf{z}(n)$ is additive white Gaussian noise (AWGN) with zero-mean and variance $\sg_A^2$.
We can rewrite (\ref{eq:3}) in the matrix form as
\begin{equation}
\mathbf{Y}_A = \sqrt{P_B} \mathbf{h}_B \mathbf{b}^T + \mathbf{Z}
\end{equation}
where $\mathbf{Y}_A \triangleq [\mathbf{y}_A(1),\ldots,\mathbf{y}_A(N)]$ and $\mathbf{b} \triangleq [b(1),\ldots,b(N)]^T$.
Knowing the pilot signals $\mathbf{b}$, Alice can easily estimate the channel $\mathbf{h}_B$ by the least-squares approach
\begin{eqnarray}
\mathbf{\widehat{h}}_B &=& \frac{1}{N} \mathbf{Y}_A \mathbf{b} \non \\
&=& \sqrt{P_B} \mathbf{h}_B + \frac{1}{N} \mathbf{Z} \mathbf{b}. \label{eq:CCE}
\end{eqnarray}
Then, Alice utilizes $\mathbf{\widehat{h}}_B$ to design the beamformer $\mathbf{w}$ for the downlink data transmission
\begin{equation}
\mathbf{w} = \frac{\mathbf{\widehat{h}}_B}{\|\mathbf{\widehat{h}}_B\|}
\label{eq:beamformer}
\end{equation}
which can provide Bob with maximum receive SNR.
With this beamformer,  the received signal at Bob in the $n$th time slot  is
\begin{equation}
y_B(n) = \sqrt{P_A}  \mathbf{h}_B^H \mathbf{w} s(n) + v(n)
\label{eq:received signal Bob}
\end{equation}
where $P_A$ is the data transmission power of Alice, $s(n)$ is the transmitted data, $\mathbb{E}\{|s(n)|^2\} = 1$ and $v(n)$ is AWGN at Bob with zero-mean and variance $\sg_{B}^2$.

Here we use the average SNR at Bob to evaluate the performance of legitimate transmission
\begin{eqnarray}
\mathrm{\overline{SNR}}_B & = &  \mathbb{E}_{\mathbf{h}_B} \left\{ \frac{P_A \| \mathbf{h}_B^H \mathbf{w} \|^2}{\sg_B^2} \right\}  \non \\
&=& \frac{P_A\al_B}{\sg_B^2} \cdot \frac{MNP_B\al_B + \sg_A^2}{NP_B\al_B + \sg_A^2}
\label{eq:SNR B no spoof}
\end{eqnarray}
whose proof is offered in the Appendix.
By using a beamformer $\mathbf{w}$ based on the estimated channel $\widehat{\mathbf{h}}_B$, the performance of data transmission can be improved dramatically.

\subsection{Impact of Spoofing}

If Eve conducts the pilot spoofing attack by sending the same pilot signals at the same time, the received signal at Alice will become
\begin{equation}
\mathbf{y}_A(n) = \sqrt{P_B} \mathbf{h}_B b(n) + \sqrt{P_E} \mathbf{h}_E b(n) + \mathbf{z}(n), n=1,\ldots,N.
\end{equation}
After rewriting it in the matrix form
\begin{equation}
\mathbf{Y}_A = (\sqrt{P_B} \mathbf{h}_B + \sqrt {P_E} \mathbf{h}_E)  \mathbf{b}^T + \mathbf{Z},
\end{equation}
the estimated channel $\mathbf{h}_B$ becomes
\begin{eqnarray}
\mathbf{\widehat{h}}_B & = & \frac{1}{N} \mathbf{Y}_A \mathbf{b}   \non \\
&=& \sqrt{P_B} \mathbf{h}_B + \sqrt{P_E} \mathbf{h}_E + \frac{1}{N} \mathbf{Z} \mathbf{b}.
\end{eqnarray}
In general, $\mathbf{h}_E$ and $\mathbf{h}_B$ are not in the same direction, i.e. $\mathbf{h}_E \neq \al \mathbf{h}_B$ ($\al \neq 0$). If Alice still uses this contaminated/spoofed $\mathbf{\widehat{h}}_B$ to design the beamformer as in (\ref{eq:beamformer}), the direction of the signals transmitted by Alice will turn towards the direction of Eve.

Specifically, when Bob is under the spoofing attack, according to the received signal at Bob in (\ref{eq:received signal Bob}), the average SNR at Bob has the expression as
\begin{eqnarray}
\mathrm{\overline{SNR}}_B & = & \mathbb{E}_{\mathbf{h}_B, \mathbf{h}_E} \left\{ \frac{P_A \| \mathbf{h}_B^H \mathbf{w} \|^2}{\sg_B^2} \right\} \non \\
&=& \frac{P_A\al_B}{\sg_B^2} \cdot \frac{MNP_B\al_B + NP_E\al_E + \sg_A^2}{NP_B\al_B + NP_E\al_E + \sg_A^2}
\label{eq:SNR B spoof}
\end{eqnarray}
\nid whose proof is offered in the Appendix. Similarly,  the received signal at Eve in the $n$th time slot  is
\begin{equation}
y_E(n) = \sqrt{P_A} \mathbf{h}_E^H \mathbf{w}  s(n) + v'(n)
\label{eq:received signal Eve}
\end{equation}
\nid where  $v'(n)$ is AWGN at Eve with zero-mean and variance $\sg_E^2$. The average SNR at Eve has the expression as
\begin{eqnarray}
\mathrm{\overline{SNR}}_E &=& \mathbb{E}_{\mathbf{h}_B, \mathbf{h}_E} \left\{ \frac{P_A \| \mathbf{h}_E^H \mathbf{w} \|^2}{\sg_E^2} \right\} \non \\
&=& \frac{P_A\al_E}{\sg_E^2} \cdot \frac{MNP_E\al_E + NP_B\al_B + \sg_A^2}{NP_E\al_E + NP_B\al_B + \sg_A^2}
\label{eq:SNR E spoof}
\end{eqnarray}
\nid whose proof can be found in the Appendix.
An obvious conclusion is that with spoofing power $P_E$ increasing, the $\mathrm{\overline{SNR}}_B$ will decrease while the $\mathrm{\overline{SNR}}_E$ will increase.
In other words, by launching the spoofing attack, Eve can manipulate the channel estimation result and benefit from degrading legitimate transmission performance as well as enhancing eavesdropping.

\begin{center}
\begin{figure}[t]
\includegraphics[width=3.6 in]{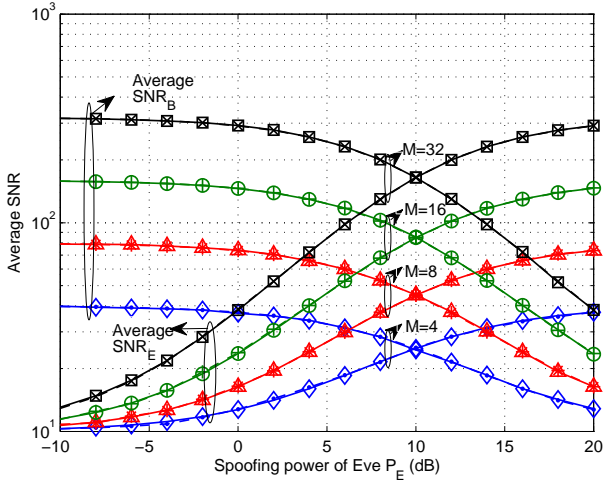} \vspace{-0.4 cm}
\caption{The average SNRs at Bob and Eve versus the spoofing power of Eve. The solid lines represent the theoretical results and the dashed lines represent the simulation results.}\vspace*{-0.0 cm} \label{fg:SNR}

\end{figure}
\end{center}

To illustrate the impact due to the pilot spoofing attack, we carry out a simulation as  shown in Fig. \ref{fg:SNR}. In this simulation, we set $\sg_A^2=\sg_B^2=\sg_E^2=1$, $\al_B=\al_E=1$, $P_B=10\mathrm{dB}$.
The number of pilot bits is $N=100$ and the number of antennas at Alice is set as $M=4, 8, 16,$ and $32$, respectively. Both theoretical results  and simulation results are included.

\vspace{-0.0 cm}

\vspace{-0.9 cm}

It can be observed from Fig. \ref{fg:SNR} that the theoretical results (the solid lines) perfectly match the simulation results (the dashed lines).
When $P_E$ increases, $\mathrm{\overline{SNR}}_B$ decreases and the legitimate transmission performance is degraded. More importantly,  $\mathrm{\overline{SNR}}_E$ becomes larger as $P_E$ increasing,  which implies that Eve can successfully spoof Alice and eavesdrop the transmitted information.

While the consequence of pilot spoofing attack is unbearable, it is of great importance that Alice and Bob identify this type of attack and then take actions if the spoofing attack is detected. In the following section, we will present our RTA pilot spoofing detection scheme and provide the theoretical performance analysis of it.

\section{Random-Training-Assisted Pilot Spoofing Detection}
\label{sec: ourscheme}

\vspace{0.0 cm}
\begin{center}
\begin{figure}[t]
\begin{center}
\vspace*{0.5 cm}
\includegraphics[width=3.0 in]{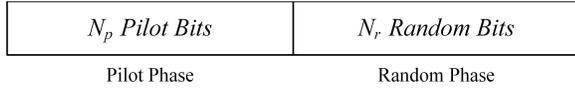} \vspace{-0.0 cm}
\caption{Structure of pilot and random training signals.} \label{fg:sgstructure}\vspace{-0.0 cm}
\end{center}
\end{figure}
\end{center}
\vspace{-0.0 cm}


Considering the serious impact due to the pilot spoofing attack discussed in the previous section, in this section we will develop a novel pilot spoofing detector which is based solely on uplink training with the assist of random training. In particular, we develop a new training mechanism
by adding a \textit{Random Phase} following the conventional
\textit{Pilot Phase}. For uplink channel estimation,  Bob first transmits  $N_p$ pre-designed pilot bits during the pilot phase, then sends $N_r$ random bits during the random phase, as illustrated in Fig. \ref{fg:sgstructure}. While the pilot bits are publicly known to Alice, Bob, and Eve, the random bits are randomly generated by Bob and unknown to neither Alice nor Eve. In the following, we will introduce our proposed spoofing detection algorithm in detail for \textit{Pilot Phase} and \textit{Random Phase}, respectively. For the convenience of developing our scheme, we define two hypothesises: $\mathcal{H}_0$ denotes that there is no pilot spoofing attack; $\mathcal{H}_1$ denotes that Alice and Bob are under pilot spoofing attack.

\vspace*{0.5 cm}

\subsection{Pilot Phase}
During the pilot phase, Bob transmits $N_p$ pilot bits, $b_p(n)\in \{ \pm 1 \}$, $n=1,\ldots, N_p$, and Eve may transmit the same pilot bit sequence as Bob since the pilot bit sequence is publicly known.

Under the hypothesis of $\mathcal{H}_0$, the received signal at Alice in the $n$th time slot is
\begin{equation}
\mathbf{y}_p(n) = \sqrt{P_B} \mathbf{h}_B b_p(n) + \mathbf{u}_p(n), n=1, \ldots ,N_p,
\end{equation}
where the AWGN $\mathbf{u}_p(n) \sim \mathcal{CN}(\mathbf{0}, \sg_A^2 \mathbf{I}_M)$.
Rewrite it in the matrix form
\begin{equation}
\mathbf{Y}_p = \sqrt{P_B} \mathbf{h}_B \mathbf{b}_p^T + \mathbf{U}_p
\label{eq:Alice H0 pilot phase}
\end{equation}
where $\mathbf{Y}_p \triangleq [\mathbf{y}_p(1),\mathbf{y}_p(2), \ldots ,\mathbf{y}_p(N_p)] \in \mathbb{C}^{M \times N_p}$.
Then, the channel estimation result $\mathbf{\widehat{h}}_{Bp} = \frac{1}{N_p} \mathbf{Y}_p \mathbf{b}_p$ under $\mathcal{H}_0$ is
\begin{eqnarray}
\mathcal{H}_0 : \mathbf{\widehat{h}}_{Bp} &=& \sqrt{P_B} \mathbf{h}_B + \frac{1}{N_p} \mathbf{U}_p \mathbf{b}_p  \non \\
&=& \sqrt{P_B} \mathbf{h}_B + \triangle\mathbf{h}_{Bp}^{\mathcal{H}_0}
\label{eq:channel est H0 pilot}
\end{eqnarray}
where $\triangle\mathbf{h}_{Bp}^{\mathcal{H}_0} \triangleq \frac{1}{N_p} \mathbf{U}_p \mathbf{b}_p$ is the estimation error and by central limited theorem
\begin{equation}
\triangle\mathbf{h}_{Bp}^{\mathcal{H}_0}   \sim \mathcal{CN}(\mathbf{0}, \frac{\sg_A^2}{N_p}\mathbf{I}_M).
\label{eq:pilot H0 distribution}
\end{equation}

Under the hypothesis of $ \mathcal{H}_1$, the received signal at Alice in the $n$th time slot is
\begin{equation}
\mathbf{y}_p(n) \hspace{-0.1 cm} = \hspace{-0.1 cm} \sqrt{P_B} \mathbf{h}_B b_p(n) + \sqrt{P_E} \mathbf{h}_E b_p(n)  +  \mathbf{u}_p(n), n=1,\ldots,N_p.
\end{equation}
Similarly, we can rewrite it in the matrix form
\begin{equation}
\mathbf{Y}_p  =  \sqrt{P_B} \mathbf{h}_B \mathbf{b}_p^T + \sqrt{P_E} \mathbf{h}_E \mathbf{b}_p^T + \mathbf{U}_p
\label{eq:Alice H1 pilot phase}
\end{equation}
\nid and the channel estimation result under $\mathcal{H}_1$ is

\begin{eqnarray}
\mathcal{H}_1 : \mathbf{\widehat{h}}_{Bp} & = & \sqrt{P_B} \mathbf{h}_B + \sqrt{P_E} \mathbf{h}_E + \frac{1}{N_p} \mathbf{U}_p \mathbf{b}_p  \non \\
&=& \sqrt{P_B} \mathbf{h}_B + \triangle\mathbf{h}_{Bp}^{\mathcal{H}_1}
\label{eq:channel est H1 pilot}
\end{eqnarray}
where $\triangle\mathbf{h}_{Bp}^{\mathcal{H}_1} \triangleq \sqrt{P_E} \mathbf{h}_E + \frac{1}{N_p} \mathbf{U}_p \mathbf{b}_p$ is the estimation error.
If we consider $\sqrt{P_E} \mathbf{h}_E$ as random variable with variance $P_E \al_E \mathbf{I}_M$, then
\begin{equation}
\triangle\mathbf{h}_{Bp}^{\mathcal{H}_1} \sim \mathcal{CN}(\mathbf{0}, (\frac{\sg_A^2}{N_p}+P_E \al_E)\mathbf{I}_M).
\label{eq:pilot H1 distribution}
\end{equation}

\nid Comparing to (\ref{eq:channel est H0 pilot}), we can observe that the channel estimation (\ref{eq:channel est H1 pilot}) under $\mathcal{H}_1$ is biased and turn towards  $\mathbf{h}_E$ due to the attack launched by Eve.

\subsection{Random Phase}
During the random phase, Bob transmits $N_r$ random bits $b_r(n) \in \{ \pm 1 \}$ while Eve sends nothing since she does not have any knowledge of the random bits generated by Bob.
The received signal at Alice in the $n$th time slot is
\begin{equation}
\mathbf{y}_r(n) = \sqrt{P_B} \mathbf{h}_B b_r(n) + \mathbf{u}_r(n), n=1,\ldots ,N_r,
\end{equation}
\nid where $\mathbf{u}_r(n) \sim \mathcal{CN}(\mathbf{0}, \sg_A^2 \mathbf{I}_M)$ is the AWGN. Rewrite it in the matrix form
\begin{equation}
\mathbf{Y}_r = \sqrt{P_B} \mathbf{h}_B \mathbf{b}_r^T + \mathbf{U}_r
\label{eq:Alice H0 random phase}
\end{equation}
where $\mathbf{Y}_r \triangleq [\mathbf{y}_r(1),\mathbf{y}_r(2), \ldots ,\mathbf{y}_r(N_r)] \in  \mathbb{C}^{M \times N_r}$.

Without knowing the random bits $\mathbf{b}_r$, we propose for Alice to adopt  the iterative least-squares (ILS) algorithm \cite{Li 15}, \cite{Li 16} to blindly estimate $\mathbf{h}_B$. The ILS algorithm is a blind joint symbol detection and channel estimation algorithm without availability of  pilot signals,  which has not only high accuracy but also low complexity.  The basic idea of the ILS algorithm is to iteratively execute the symbol detection and the channel estimation until convergence. In order to obtain a reliable solution, we use the estimated channel $\widehat{\mathbf{h}}_{Bp}$ acquired by the pilot bits as the initialization of the ILS algorithm. Due to the space limitation, we would not review the ILS algorithm in detail but simply summarize it in Table \ref{table:ILS}. Superscript $d$ denotes iteration index.

\begin{center}
\begin{table}[t]  \vspace{0.3 cm}\caption{
Iterative  Least-Squares (ILS) Algorithm} \vspace{-0.2
cm}\begin{center}
\begin{tabular}{l}
\hline \hline \vspace{-0.2 cm}\\
\hspace{-0.2 cm}  \textbf{Input:} $\widehat{\mathbf{h}}_{Bp}$, $\mathbf{Y}_r$.  \\
\hspace{-0.2 cm}  \textbf{Output:} $\mathbf{{\widehat{h}}}_{Br}$. \\
 \hspace{-0.2 cm} 1) $d=0$; initialize $\mathbf{\widehat{h}}_{Br}^{(0)}  = \widehat{\mathbf{h}}_{Bp}$.\\
\hspace{-0.2 cm} 2) $d=d+1$; \\
\hspace{0.2 cm}  \begin{small}  $ \mathbf{{\widehat{b}}}_r^{(d)}= \mathrm{sgn}\left\{
\mathfrak{Re}\left[(\mathbf{\widehat{h}}_{Br}^{(d-1)})^H\mathbf{Y}_r
  \right]\right\}$; \end{small} \\
\hspace{0.3 cm}  \begin{small}$\mathbf{{\widehat{h}}}_{Br}^{(d)}
=\frac{1}{N_r}\mathbf{Y}_r \mathbf{\widehat{b}}_r^{(d)} $.  \end{small} \\
\hspace{-0.2 cm} 3) Repeat Step 2 until
$ \mathbf{{\widehat{h}}}_{Br}^{(d)} = \mathbf{{\widehat{h}}}_{Br}^{(d-1)} $.\vspace{0.1 cm} \\
\hline
\end{tabular}\label{table:ILS}
\end{center}
\end{table}
\end{center}

\vspace{-0.3 cm}

If the channel estimation result $\mathbf{{\widehat{h}}}_{Br}$ acquired in the random phase using ILS algorithm is accurate, then for both $\mathcal{H}_0$ and $\mathcal{H}_1$, the channel estimation will be $\mathbf{\widehat{h}}_{Br} = \frac{1}{N_r} \mathbf{Y}_r \mathbf{b}_r^T$ and we have
\begin{eqnarray}
\mathcal{H}_0 , \mathcal{H}_1 : \mathbf{\widehat{h}}_{Br} &=& \sqrt{P_B} \mathbf{h}_B + \frac{1}{N_r} \mathbf{U}_r \mathbf{b}_r  \non \\
&=& \sqrt{P_B} \mathbf{h}_B + \triangle\mathbf{h}_{Br} ^{{\mathcal{H}_0},{\mathcal{H}_1}}
\label{eq:channel est random}
\end{eqnarray}
\nid where $\triangle\mathbf{h}_{Br}^{{\mathcal{H}_0},{\mathcal{H}_1}} \triangleq \frac{1}{N_r} \mathbf{U}_r \mathbf{b}_r^T$ is the estimation error and
\begin{equation}
\triangle\mathbf{h}_{Br}^{{\mathcal{H}_0},{\mathcal{H}_1}} \sim \mathcal{CN}(\mathbf{0}, \frac{\sg_A^2}{N_r}\mathbf{I}_M).
\label{eq:random H0H1 distribution}
\end{equation}

%
%

While the channel estimation $ \mathbf{\widehat{h}}_{Br}$ for the random phase is reliable under both $\mathcal{H}_0$ and $\mathcal{H}_1$, we recall that $ \mathbf{\widehat{h}}_{Bp}$ for the pilot phase is reliable under $\mathcal{H}_0$ but biased under $\mathcal{H}_1$. In the following, we will develop our spoofing attack detection algorithm by examining the difference
of the estimated legitimate channels during these two phases.

%

\subsection{Spoofing Attack Detector}

According to (\ref{eq:channel est H0 pilot}), (\ref{eq:channel est H1 pilot}) and (\ref{eq:channel est random}), the difference between $\mathbf{\widehat{h}}_{Bp}$ and $\mathbf{\widehat{h}}_{Br}$ under $\mathcal{H}_0$ and $\mathcal{H}_1$ has the expression as
\begin{eqnarray}
\hspace*{-0.4 cm}\mathcal{H}_0 \hspace{-0.3 cm} & : & \hspace*{-0.3 cm} \mathbf{\widehat{h}}_{Bp} - \mathbf{\widehat{h}}_{Br}   = \triangle\mathbf{h}_{Bp}^{\mathcal{H}_0} - \triangle\mathbf{h}_{Br}^{\mathcal{H}_0} \non \\
&& \hspace{1.45 cm} = \frac{1}{N_p} \mathbf{U}_p \mathbf{b}_p  - \frac{1}{N_r} \mathbf{U}_r \mathbf{b}_r, \\
\hspace*{-0.4 cm} \mathcal{H}_1 \hspace{-0.3 cm} & : & \hspace*{-0.3 cm} \mathbf{\widehat{h}}_{Bp} - \mathbf{\widehat{h}}_{Br}   = \triangle\mathbf{h}_{Bp}^{\mathcal{H}_1} - \triangle\mathbf{h}_{Br}^{\mathcal{H}_1} \non \\ && \hspace{1.5 cm}
=  \sqrt{P_E} \mathbf{h}_E + \frac{1}{N_p} \mathbf{U}_p \mathbf{b}_p - \frac{1}{N_r} \mathbf{U}_r \mathbf{b}_r.
\end{eqnarray}

\nid Notice that under $\mathcal{H}_0$  the difference contains just the noise terms and has relatively small value, while under $\mathcal{H}_1$ it has illegitimate channel component $\sqrt{P_E} \mathbf{h}_E$ and is  more significant.
Based on this fact, we introduce the test statistic as
\begin{equation}
T \triangleq \|\mathbf{\widehat{h}}_{Bp} -  \mathbf{\widehat{h}}_{Br}\|^2
\end{equation}
to differentiate $\mathcal{H}_0$ and $\mathcal{H}_1$, and to implement the spoofing attack detection. The hypothesis test problem of the detector is
\begin{equation}
T \underset{\mathcal{H}_1}{\overset{\mathcal{H}_0}{\lessgtr}}  \gamma
\label{eq:detector}\end{equation}
where $\gamma$ is the test threshold.

Now we attempt to provide the theoretical performance analysis of the proposed spoofing detector in (\ref{eq:detector}) and determine appropriate test threshold $\gamma$.
According to (\ref{eq:pilot H0 distribution}), (\ref{eq:pilot H1 distribution}) and (\ref{eq:random H0H1 distribution}), we have the distribution of ($\triangle \mathbf{\widehat{h}}_{Bp}^{\mathcal{H}_0} -  \triangle  \mathbf{\widehat{h}}_{Br}^{\mathcal{H}_0}$) and ($  \triangle \mathbf{\widehat{h}}_{Bp}^{\mathcal{H}_1} -   \triangle \mathbf{\widehat{h}}_{Br}^{\mathcal{H}_1}$) as
\begin{align}
\mathcal{H}_0 &:    \triangle\mathbf{\widehat{h}}_{Bp}^{\mathcal{H}_0} -  \triangle\mathbf{\widehat{h}}_{Br}^{\mathcal{H}_0}   \sim \mathcal{CN} (\mathbf{0},  (\frac{\sg_A^2}{N_p}+\frac{\sg_A^2}{N_r}  )\mathbf{I}_M ), \\
\mathcal{H}_1 &:   \triangle\mathbf{\widehat{h}}_{Bp}^{\mathcal{H}_1} -  \triangle  \mathbf{\widehat{h}}_{Br}^{\mathcal{H}_1}  \sim \mathcal{CN} (\mathbf{0}, (\frac{\sg_A^2}{N_p}+\frac{\sg_A^2}{N_r}+P_E\bt_E  )\mathbf{I}_M ).
\end{align}

Therefore, under $\mathcal{H}_0$,  the distribution of $T = \| \triangle\mathbf{\widehat{h}}_{Bp}^{\mathcal{H}_0} -  \triangle\mathbf{\widehat{h}}_{Br}^{\mathcal{H}_0} \|^2$ is chi-square multiplying a coefficient $C_0$:
\begin{eqnarray}
&T = C_0T_0', \\
&T_0' = \left\| \frac{ \triangle\mathbf{\widehat{h}}_{Bp}^{\mathcal{H}_0} -  \triangle\mathbf{\widehat{h}}_{Br}^{\mathcal{H}_0} }{C_0} \right\|^2 \sim \chi_{2M}^2, \\
&C_0 = \frac{1}{2}(\frac{\sg_A^2}{N_p}+\frac{\sg_A^2}{N_r}),
\end{eqnarray}
\nid where $\chi_{2M}^2$ denotes the chi-squared distribution with degrees of freedom $2M$.

Similarly, under $\mathcal{H}_1$, the distribution of $T = \| \triangle\mathbf{\widehat{h}}_{Bp}^{\mathcal{H}_1} -  \triangle\mathbf{\widehat{h}}_{Br}^{\mathcal{H}_1} \|^2$  is
\begin{eqnarray}
&T = C_1 T_1', \\
&T_1' = \left\| \frac{\triangle\mathbf{\widehat{h}}_{Bp}^{\mathcal{H}_1} -  \triangle\mathbf{\widehat{h}}_{Br}^{\mathcal{H}_1} }{ C_1} \right\|^2 \sim \chi_{2M}^2, \\
&C_1 = \frac{1}{2}(\frac{\sg_A^2}{N_p}+\frac{\sg_A^2}{N_r}+P_E\bt_E).
\end{eqnarray}

Given a threshold $\gamma$, we can derive the probability of false alarm $P_{fa}$ as
\begin{eqnarray}
P_{fa} &=& Pr\{T > \gamma | \mathcal{H}_0\} \non \\
&=& Pr\{T' > \frac{\gamma}{C_0} | \mathcal{H}_0\} \non \\
&=& \int_{\frac{\gamma}{C_0}}^{+\infty} f(t)\mathrm{d}t \non \\
&=& 1 - F(\frac{\gamma}{C_0})
\label{eq:pfa}
\end{eqnarray}
where $f(t)$ and $F(t)$ are the probability density function (PDF) and corresponding cumulative distribution function (CDF) of $\chi_{2M}^2$, respectively.
Similarly, with the threshold $\gamma$,  the probability of detection $P_d$ of our proposed detector is
\begin{eqnarray}
P_{d} &=& Pr\{T > \gamma | \mathcal{H}_1\} \non \\
&=& Pr\{T' > \frac{\gamma}{C_1} | \mathcal{H}_1\} \non \\
&=& \int_{\frac{\gamma}{C_1}}^{+\infty} f(t)\mathrm{d}t \non \\
&=& 1 - F(\frac{\gamma}{C_1}).
\label{eq:pd}
\end{eqnarray}

Given a $P_{fa}$ (or $P_{d}$), the corresponding threshold $\gamma$ can be obtained by calculating the inverse function of $F(\frac{\gamma}{C_0})$ (or $F(\frac{\gamma}{C_1})$).

\subsection{Eve Attacks During Random Phase}
Under the fore-mentioned assumption, Eve transmits nothing during the random phase. However, obviously Eve can also attack and transmit some signals  during this phase, for example Eve may transmit random bits or random Gaussian noise. Here, we briefly explain how our detector works under this case.

In this case, Bob transmits $N_r$ random bits $b_r(n)$ and Eve transmits $N_r$ different random signal $b_{Er}(n)$ at the same time during the random phase.
The received signal at Alice in the $n$th time slot is
\begin{equation}
\mathbf{y}_r(n) = \sqrt{P_B} \mathbf{h}_B b_r(n) + \sqrt{P_{Er}} \mathbf{h}_E b_{Er}(n) + \mathbf{u}_r(n)
\end{equation}
where $\sqrt{P_{Er}}$ is the power of the spoofing attack of Eve during the random phase. Again, we rewrite it in the matrix form as
\begin{equation}
\mathbf{Y}_r = \sqrt{P_B} \mathbf{h}_B \mathbf{b}_r^T + \sqrt{P_{Er}} \mathbf{h}_E \mathbf{b}_{Er}^T + \mathbf{U}_r.
\end{equation}

Due to this attack, the  legitimate channel estimation and the random bit detection returned by the ILS algorithm may be not accurate.
Specifically, let $\widehat{\mathbf{b}}_r$ be the detected random bit vector returned by ILS. Then, The channel estimation result $\mathbf{\widehat{h}}_{Br}$ for $\mathcal{H}_1$ can be expressed as below
\bea
 \mathbf{\widehat{h}}_{Br} \hspace{-0.2 cm} & = & \hspace{-0.2 cm} \frac{1}{N_r}\sqrt{P_B} \mathbf{h}_B \mathbf{b}_r^T \widehat{\mathbf{b}}_r  + \frac{1}{N_r}\sqrt{P_{Er}} \mathbf{h}_E \mathbf{b}_{Er}^T\widehat{\mathbf{b}}_r  + \frac{1}{N_r}\mathbf{U}_r\widehat{\mathbf{b}}_r \non \\
\hspace{-0.2 cm} & = & \hspace{-0.2 cm}   \mu \sqrt{P_B} \mathbf{h}_B + \nu \sqrt{P_{Er}} \mathbf{h}_E   + \frac{1}{N_r}\mathbf{U}_r \widehat{\mathbf{b}}_r
\label{eq:channel est random case 1}
\eea
where $  \mu \triangleq \frac{1}{N_r} \mathbf{b}_r^T  \widehat{\mathbf{b}}_r $, $|\mu| \leq 1 $ and the equality holds if and only if $\mathbf{b}_r = \pm \widehat{\mathbf{b}}_r $, i.e. the detection of $\mathbf{b}_r$ is correct.
  Similarly,  $\nu \triangleq \frac{1}{N_r} \mathbf{b}_{Er}^T \widehat{\mathbf{b}}_r  $, $|\nu| \leq 1 $ and the equality holds if and only if $\mathbf{b}_{Er} =  \pm \widehat{\mathbf{b}}_r $, in such case the ILS total failed and return the detection of $\mathbf{b}_{Er} $ instead.

Then, we can rewrite the difference between $\mathbf{\widehat{h}}_{Bp}$ and $\mathbf{\widehat{h}}_{Br}$ as
\begin{eqnarray}
\mathcal{H}_0 : \mathbf{\widehat{h}}_{Bp} - \mathbf{\widehat{h}}_{Br} \hspace{-0.1 cm} &=& \hspace{-0.1 cm} \frac{1}{N_p} \mathbf{U}_p \mathbf{b}_p  - \frac{1}{N_r} \mathbf{U}_r \mathbf{b}_r \non \\
\mathcal{H}_1 : \mathbf{\widehat{h}}_{Bp} - \mathbf{\widehat{h}}_{Br}
\hspace{-0.1 cm} &=& \hspace{-0.1 cm}  \frac{1}{N_p} \mathbf{U}_p \mathbf{b}_p   -   \frac{1}{N_r} \mathbf{U}_r \widehat{\mathbf{b}}_r \non \\   \hspace{-0.1 cm} & & \hspace{-0.1 cm} +   (1 - \mu) \sqrt{P_B} \mathbf{h}_B + (1 - \nu) \sqrt{P_{Er}} \mathbf{h}_E . \non
\end{eqnarray}


Since $\mathbf{b}_r$ and $\mathbf{b}_{Er}$ are independent identically distributed (i.i.d.), $\mathbf{b}_r$ and $\mathbf{b}_{Er}$ are quasi-orthogonal with sufficient length of random bits.
Due to this property, no matter what result of $\widehat{\mathbf{b}}_r$ returned by the ILS algorithm, $|\nu|$ and $|\mu|$ cannot simultaneously have large value close to 1. Therefore, under Eve's attack during the random phase, the residual of the channel estimation is still significant and our detector is functional under this case.


\section{Channel Estimation Enhancement}
\label{sec: enhencement}

%

If the spoofing detector indicates absence of any spoofing attack, the channel estimations $\mathbf{\widehat{h}}_{Bp}$ and $\mathbf{\widehat{h}}_{Br}$ are both reliable for Alice to design the beamformer.
However, it is well known that larger number of pilot bits can provide more  accurate channel estimation. If we have the knowledge of the random bits and use them together with the pilot bits to estimate the channel, the accuracy  will be further improved. Motivated by this fact, in this section we introduce a computationally efficient channel estimation enhancement algorithm.

We first use $\mathbf{\widehat{h}}_{Bp}$ obtained by the pilot bits to detect the random bits
\be  \mathbf{{\widehat{b}}}_r = \mathrm{sgn}\left\{
\mathfrak{Re}\left[\mathbf{\widehat{h}}_{Bp}^H\mathbf{Y}_r
  \right]\right\}. \label{eq:aaa}  \ee

\nid After combining the pilot and random bits as $\widehat{\mathbf{b}}_c = [\mathbf{b}_p^T, (\mathbf{{\widehat{b}}}_r)^T]^T$, we then use $\widehat{\mathbf{b}}_c$ to estimate the channel based on the concatenated received data $\mathbf{Y}_c = [\mathbf{Y}_p , \mathbf{Y}_r] $ from pilot and random phases:
\be \mathbf{{\widehat{h}}}_{Bc} =\frac{1}{(N_p + N_r)}\mathbf{Y}_c \mathbf{\widehat{b}}_c . \label{eq:bbb}  \ee

\nid To further improve the accuracy, we iteratively execute the random bits detection (\ref{eq:aaa}) and the channel estimation  (\ref{eq:bbb})  until convergence. The proposed ILS-based channel estimation enhancement algorithm is summarized in Table \ref{table:CEE}. Experimentally, the number of executed iteration is $2\sim4$ in general.  The simulation results show that this simple channel estimation enhancement can significantly improve the performance comparing with using only $N_p$ pilot bits and can approach the performance benchmark which uses ($N_p + N_r$) pilot bits.

\begin{center}
\begin{table}[!t]  \vspace{0.3 cm}\caption{ILS-based Channel Estimation Enhancement} \vspace{-0.2
cm}\begin{center}
\begin{tabular}{l}
\hline \hline \vspace{-0.2 cm}\\
\hspace{-0.2 cm}  \textbf{Input:} $\widehat{\mathbf{h}}_{Bp}$, $\mathbf{Y}_p$, $\mathbf{Y}_r$.  \\
\hspace{-0.2 cm}  \textbf{Output:} $\mathbf{{\widehat{h}}}_{Bc}$. \\
 \hspace{-0.2 cm} 1) $d=0$; $\mathbf{Y}_c = [\mathbf{Y}_p  ,  \mathbf{Y}_r]$. \\ \hspace{-0.2 cm} 2) Initialize $\mathbf{\widehat{h}}_{Bc}^{(0)} = \widehat{\mathbf{h}}_{Bp}$.\\
\hspace{-0.2 cm} 3) $d=d+1$; \\
\hspace{0.2 cm} $ \mathbf{{\widehat{b}}}_r^{(d)}= \mathrm{sgn}\left\{
\mathfrak{Re}\left[(\mathbf{\widehat{h}}_{Bc}^{(d-1)})^H\mathbf{Y}_r
  \right]\right\}$; \\
 \hspace{0.2 cm}   $\widehat{\mathbf{b}}_c^{(d)} = [\mathbf{b}_p^T, (\mathbf{{\widehat{b}}}_r^{(d)})^T]^T$;
   \\ 
\hspace{0.2 cm} $\mathbf{{\widehat{h}}}_{Bc}^{(d)}
=\frac{1}{(N_p + N_r)}\mathbf{Y}_c\mathbf{\widehat{b}}_c^{(d)}$.  \\
\hspace{-0.2 cm} 4) Repeat Step 3) until
$\mathbf{{\widehat{h}}}_{Bc}^{(d)}=\mathbf{{\widehat{h}}}_{Bc}^{(d-1)}$. \vspace{0.1 cm} \\
\hline
\end{tabular}\label{table:CEE}
\end{center}
\end{table}
\end{center}

\section{Secure Transmission}
\label{sec: secrecy transmission}

If the spoofing detector indicates presence of pilot spoofing attack,
Alice can of course terminate the transmission immediately. However, more wisely, she can adopt physical layer security technology and launch the secure transmission. In this section, we introduce a secure transmission method associated with the proposed spoofing detection algorithm.

We recall that, with the presence of pilot spoofing attack, the residual $\widehat{\mathbf{h}}_{Bp} - \widehat{\mathbf{h}}_{Br} $ has a form of
\begin{eqnarray}
\widehat{\mathbf{h}}_{Bp} - \widehat{\mathbf{h}}_{Br} &=& \sqrt{P_E} \mathbf{h}_E + \frac{1}{N_p} \mathbf{U}_p \mathbf{b}_p^T - \frac{1}{N_r} \mathbf{U}_r \mathbf{b}_r^T \non \\
&=& \sqrt{P_E} \mathbf{h}_E + \widetilde{\mathbf{n}}  \label{eq:h_E}
\end{eqnarray}
where $\widetilde{\mathbf{n}} \triangleq \frac{1}{N_p} \mathbf{U}_p \mathbf{b}_p^T - \frac{1}{N_r} \mathbf{U}_r \mathbf{b}_r^T $ is the AWGN term and $\widetilde{\mathbf{n}} \sim \mathcal{CN}(\mathbf{0}, (\frac{\sg_A^2}{N_p} + \frac{\sg_A^2}{N_r}) \mathbf{I}_M)$. Thus, (\ref{eq:h_E}) implies that the (energy-included) illegitimate channel can be estimated by
\begin{equation}
\mathbf{\widehat{h}}_E =  \widehat{\mathbf{h}}_{Bp} - \widehat{\mathbf{h}}_{Br} = \sqrt{P_E} \mathbf{h}_E + \widetilde{\mathbf{n}}  \label{eq:channel est eve}
\end{equation}

\nid which can be utilized in designing beamformer $\mathbf{w}$ and achieving the secure transmission in the physical layer security concept.

The optimal beamforming design $\mathbf{w}$ in the context of physical layer security is to maximize the secrecy rate $R_s (\mathbf{w})$
\bea R_s (\mathbf{w}) & = & \left[\mathrm{log}(1+\mathrm{SNR}_B) - \mathrm{log}(1+\mathrm{SNR}_E) \right]^+  \\
& = & \left[\mathrm{log} \left( \frac{\mathbf{w}^H (\mathbf{I} - \phi \mathbf{h}_B  \mathbf{h}_B^H)\mathbf{w}}{\mathbf{w}^H (\mathbf{I} -   \psi \mathbf{h}_E  \mathbf{h}_E^H)\mathbf{w}}  \right) \right]^+    \eea

\nid where $\phi \triangleq P_A/\sigma_B^2$ and $\psi \triangleq P_A/\sigma_E^2$. The maximal secrecy rate can be achieved by selecting the beamformer along the direction of the generalized eigenvector of the matrix pencil $(\mathbf{I} -  \phi \mathbf{h}_B  \mathbf{h}_B^H,\mathbf{I} -  \psi \mathbf{h}_E  \mathbf{h}_E^H)$ corresponding to the largest generalized eigenvalue \cite{Gaussian MIMO}. However, we should be aware that such optimal design requires the knowledge of $\mathbf{h}_E$, $P_E$ and $\sigma_E^2$ which are in general not always available for Alice\footnote{We recall (\ref{eq:channel est eve}) that  the estimated illegitimate channel $\widehat{\mathbf{h}}_E$ contains components of both unknown power $P_E$ and unknown channel $\mathbf{h}_E$. We cannot easily obtain them by decomposing $\widehat{\mathbf{h}}_E$.}. Considering this hurdle in the realistic applications, we turn to seek a suboptimal but practical secure transmission solution.

As mentioned in \cite{Gaussian MIMO}, in the high SNR regime, the optimal direction approaches zero-forcing (ZF), i.e. $\mathbf{h}_E \bot \mathbf{w} $. Therefore, for the practical purpose, we attempt to design beamformer $\mathbf{w}$ in a ZF manner, i.e.  maximizing the $\mathrm{SNR}_B = {P_A \| \mathbf{\widehat{h}}_B^H \mathbf{w} \|^2}/{\sg_B^2} $ while forcing the $\mathrm{SNR}_E =  {P_A \| \mathbf{\widehat{h}}_E^H \mathbf{w} \|^2}/{\sg_E^2} $ down to zero. This optimization problem can be formulated as
\bea \mathbf{w}^*_{ZF} = \mathrm{arg}
\underset{\mathbf{w} \in \mathbb{C}^M}{\mathrm{max}} & \hspace{-0.4 cm}  \| \mathbf{\widehat{h}}_B^H \mathbf{w} \|^2 \label{eq:ZF1} \\
\mathrm{s. \, \, t.} &   \| \mathbf{\widehat{h}}_E^H \mathbf{w} \|^2= 0 , \label{eq:ZF2}\\
&  \hspace{-0.7 cm} \| \mathbf{w} \| =1. \label{eq:ZF3}
\eea

\nid The optimal solution of the optimization problem (\ref{eq:ZF1})-(\ref{eq:ZF3}) can be easily obtained as
\be \mathbf{w}^*_{ZF} = \frac{\left( \mathbf{I} - \mathbf{\widehat{h}}_E \mathbf{\widehat{h}}_E^H/ \| \mathbf{\widehat{h}}_E \|^2 \right) \mathbf{\widehat{h}}_B  }{\left\| \left( \mathbf{I} - \mathbf{\widehat{h}}_E \mathbf{\widehat{h}}_E^H/ \| \mathbf{\widehat{h}}_E \|^2 \right) \mathbf{\widehat{h}}_B \right\|} \, . \label{eq:ZF_solution} \ee

\nid Notice that the secure beamforming design in (\ref{eq:ZF_solution}) only requires $\mathbf{\widehat{h}}_B$ and $\mathbf{\widehat{h}}_E$ which can be easily obtained as described before and therefore is very practical. The simulation results in the next Section illustrate that  our proposed ZF-based secure beamforming can significantly improve the secrecy rate and achieve performance as good as the generalized eigen-decomposition-based secure beamforming approach.

\section{Simulation Studies}
\label{sec: sim}

In this section, extensive simulations are carried out to illustrate the accuracy of our proposed spoofing detector as well as verify our theoretical performance analysis. We will demonstrate the spoofing detection performance for various numbers of pilots, power budgets used by Eve, numbers of antennas used by Alice. In all simulations, the small scale fading vectors  of the legitimate  and illegitimate channels, $\mathbf{\overline{h}}_B$ and $\mathbf{\overline{h}}_E$, are modeled to be Rayleigh fading with the channel vector comprising i.i.d. samples of a complex Gaussian random variable with zero mean and unit variance. The large scale fading coefficients are set as  $\al_B = \al_E = 1$. The AWGN powers at Alice, Bob, and Eve are fixed at $\sigma_A^2 = \sigma_B^2= \sigma_E^2 =1$. In all simulations, our proposed algorithm uses the same number of pilot  bits and random bits, i.e. $N_p = N_r$. The simulation results are derived from $10^6$ channel realizations. We will first focus on demonstrating the performance of our proposed algorithm itself, and then compare it with other state-of-the-art spoofing detectors to illustrate the advantage of our proposed algorithm.

\begin{center}
\begin{figure}[t]
\begin{center}
\includegraphics[width=3.6 in]{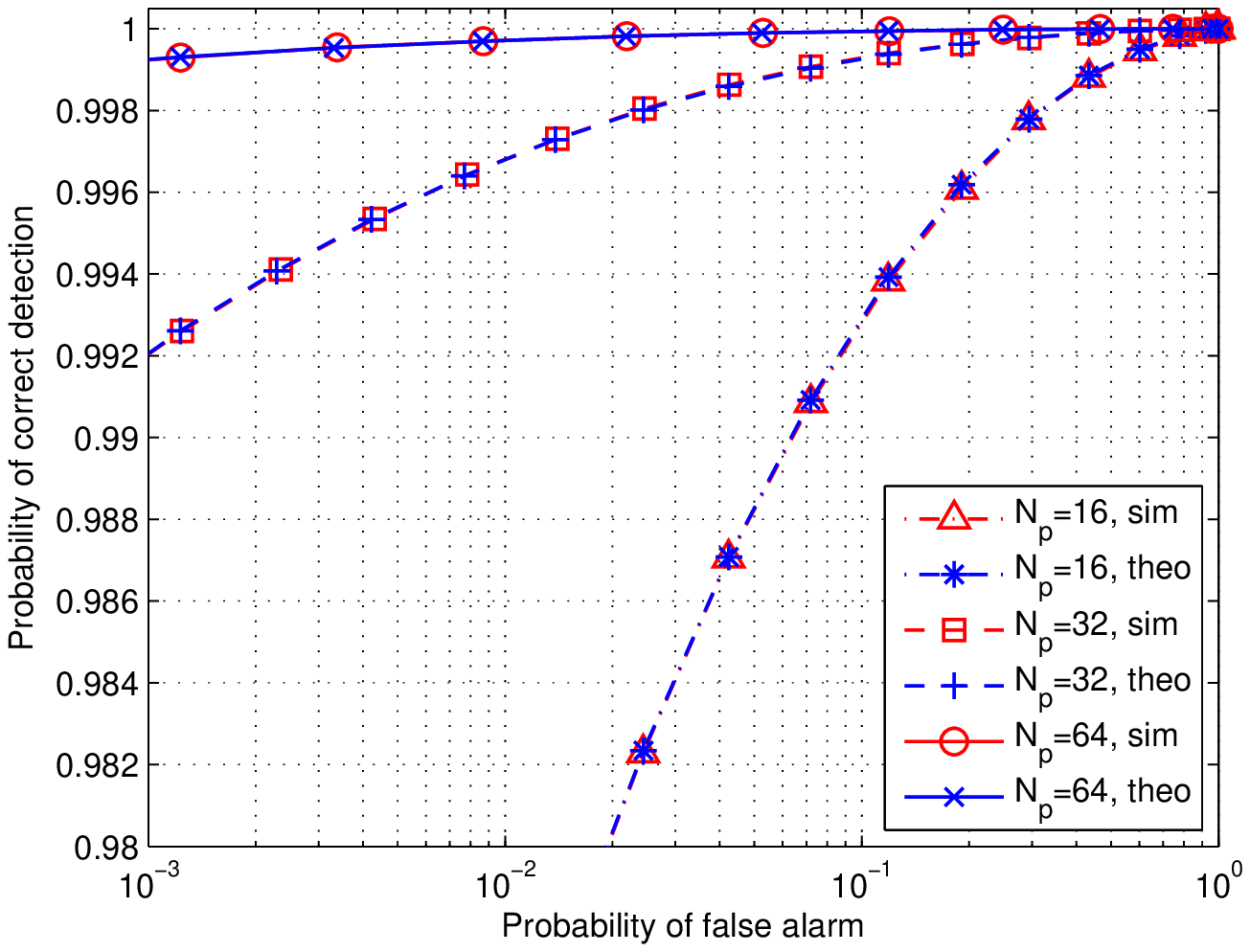} \vspace{-0.5 cm}
\caption{ROC curves of proposed spoofing detector (training power $P_B=10$, spoofing power $P_E=1$, and number of Alice's antennas $M=4$).}
\label{fg:pfavspdour}\vspace{-0.0 cm}
%
\vspace{0.3 cm}
%
\includegraphics[width=3.6 in]{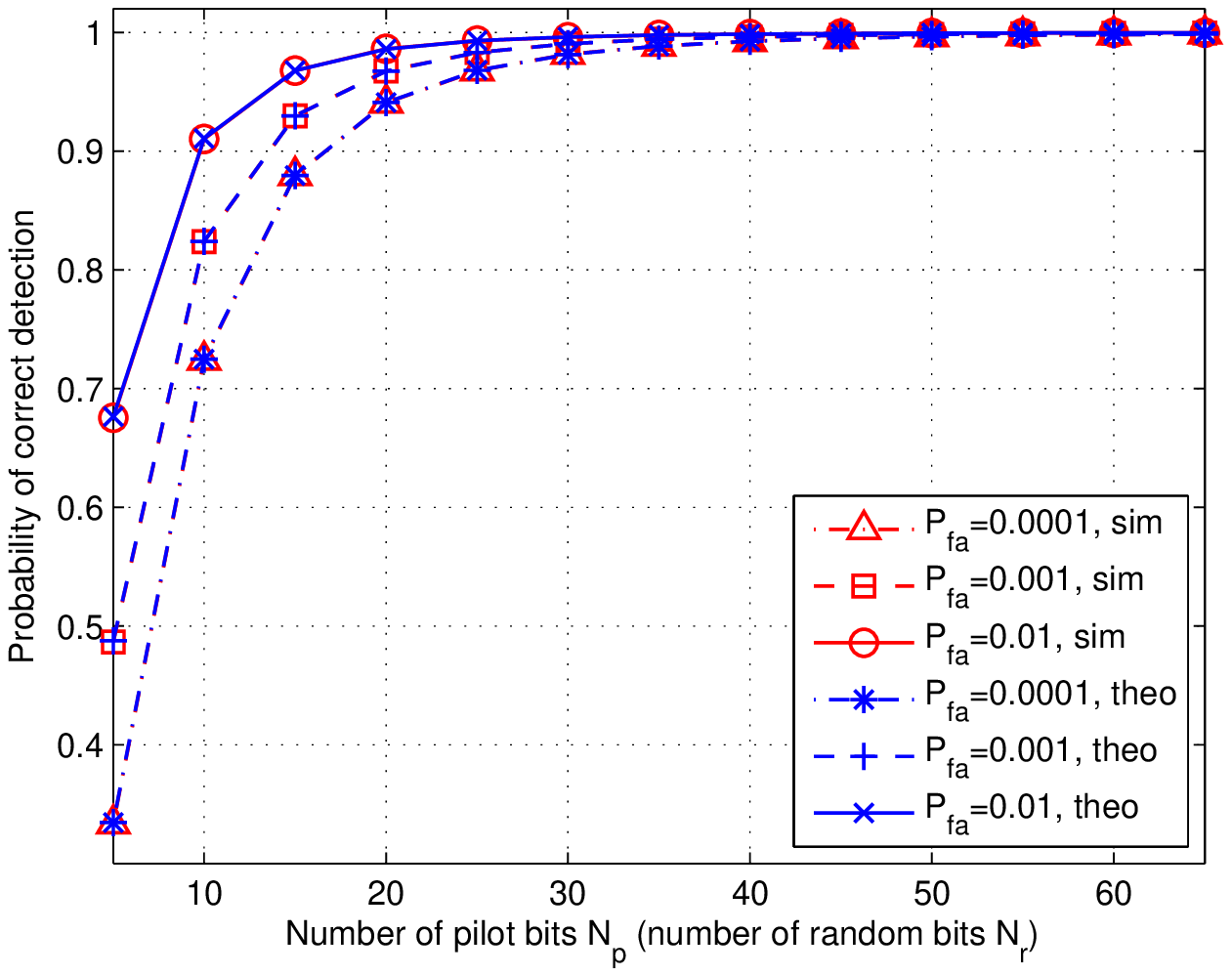} \vspace{-0.5 cm}
\caption{Probability of correct detection $P_d$ versus the number of the pilot (random) bits $N_p$ (training power $P_B=5$, spoofing power $P_E=1$, and number of Alice's antennas $M=4$).}
\label{fg:Nvspdour}\vspace{-0.0 cm}
\end{center}
\end{figure}
\end{center}

In Fig. \ref{fg:pfavspdour}, we plot the ``receiver operating characteristic'' (ROC) curves that show the probability of correct spoofing detection $P_d$ versus the probability of false alarm $P_{fa}$. In this simulation, we fix the number of Alice's antennas as $M=4$, the training power of Bob as $P_B=10$, and the spoofing power of Eve as $P_E=1$. Three ROC curves of the proposed algorithm are shown with different number of pilot bits (and random bits), $N_p=N_r=16$, $32$, and $64$. The theoretical  performances obtained by (\ref{eq:pfa}) and (\ref{eq:pd}) are also included to verify our theoretical analysis. From Fig. \ref{fg:pfavspdour} we can conclude that our proposed detector is able to accurately identify the pilot spoofing attack. Particularly, with sufficient pilot bits $N_p=64$, our proposed spoofing detection can achieve higher than $99.9\%$ accuracy with the false alarm rate at $P_{fa}=0.001$. The simulation results also perfectly match the theoretical performance analysis derived in (\ref{eq:pfa}) and (\ref{eq:pd}). Moreover, as we predicted, the larger number of pilot bits can dramatically improve detection performance.

In order to illustrate the influence of the number of pilot bits (and random bits), in Fig. \ref{fg:Nvspdour} we carry out a similar simulation to show the probability of correct detection of our proposed algorithm versus the number of pilot bits $N_p$ (number of random bits $N_r$). In this simulation, the power of pilot and random signals is $P_B=5$, spoofing power is $P_E=1$, the number of Alice's antennas is $M=4$. The results illustrate that our proposed detector is very efficient and requires only small number of pilot and random bits, for example $N_p=N_r=30$, to accurately detect the pilot spoofing attack.

\vspace{-0.0 cm}
\vspace{-0.0 cm}

\vspace{-0.4 cm}

\begin{center}
\begin{figure}[t]
\begin{center}
\includegraphics[width=3.6 in]{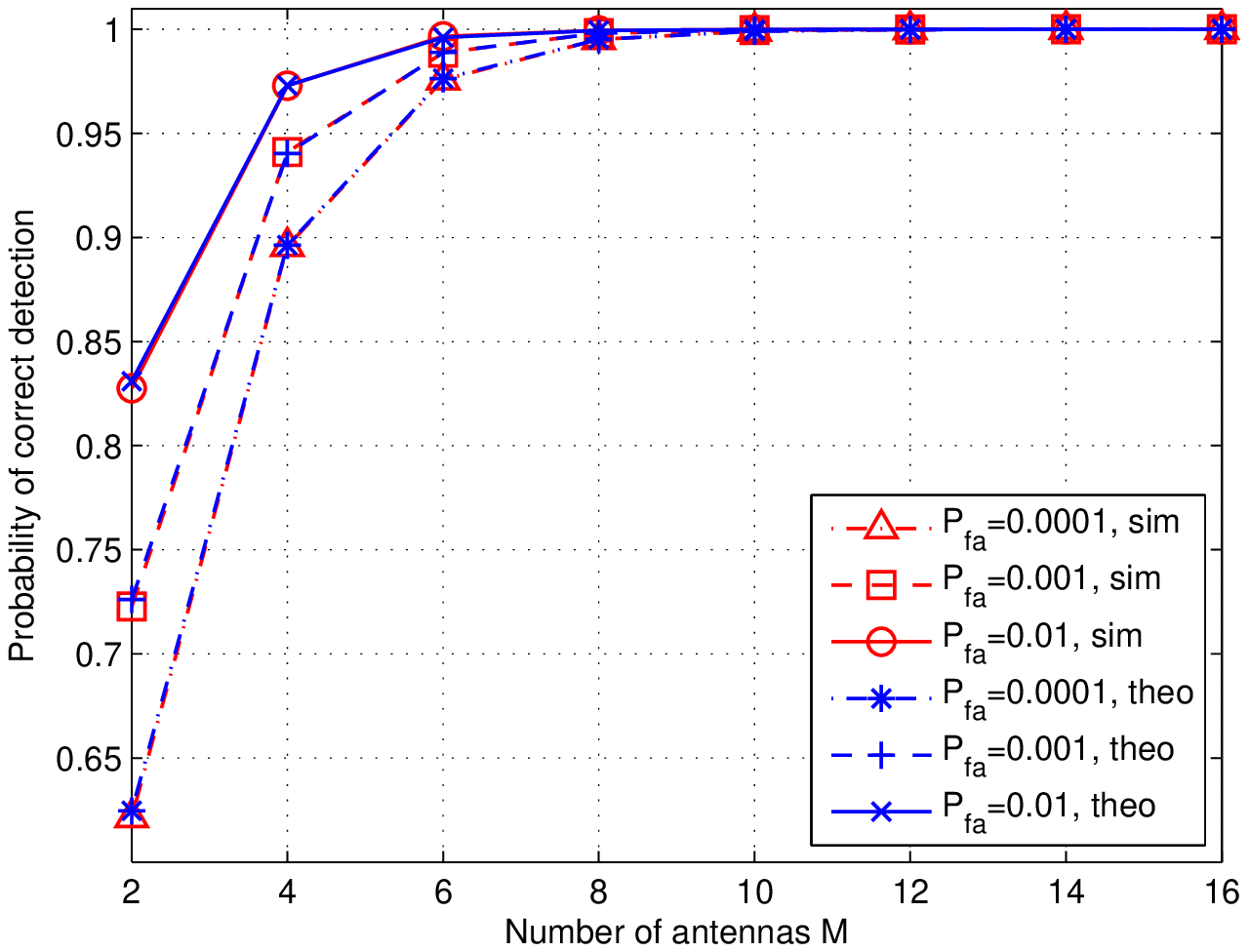} \vspace{-0.5 cm}
\caption{Probability of correct detection $P_d$ versus the number of Alice's antennas $M$ (number of pilot (random) bits $N_p=N_r=16$, training power $P_B=5$, and spoofing power $P_E=1$).}
\label{fg:Mvspdour}\vspace{-0.0 cm}
%
%
\vspace{0.3 cm}
\includegraphics[width=3.6 in]{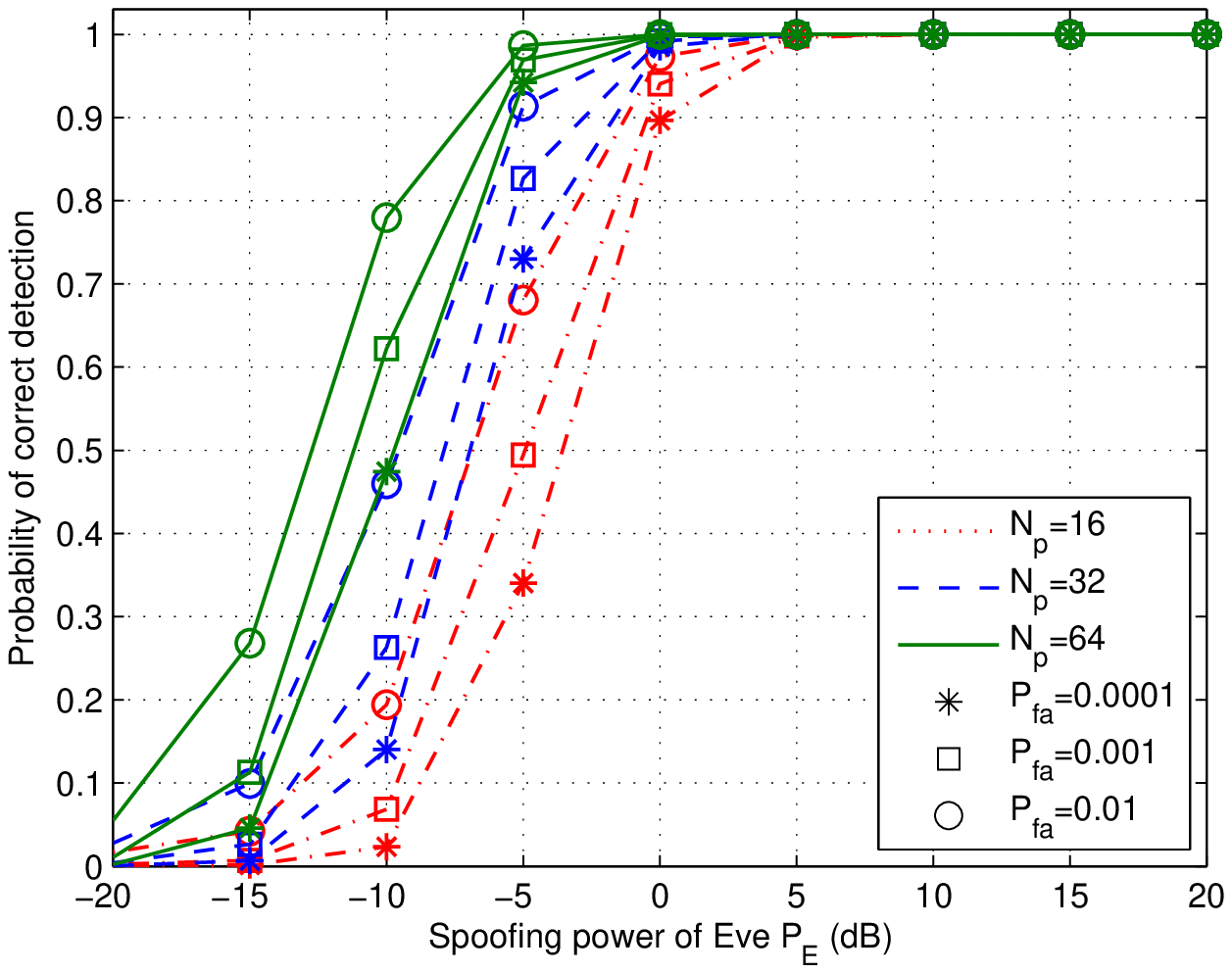} \vspace{-0.5 cm}
\caption{Probability of correct detection $P_d$ versus the spoofing power $P_E$ (training power $P_B=5$, and number of Alice's antennas $M=4$).}
\label{fg:Pevspdour}\vspace{-0.0  cm}
\end{center}
\end{figure}
\end{center}

Next, in Fig. \ref{fg:Mvspdour} we show the probability of correct detection as a function of the number of Alice's antennas $M$. Clearly, larger number of antennas, which can provide more spatial degrees of freedom and signal processing power, results in more accurate spoofing detection performance but  higher hardware cost and complexity. With the moderate number of antennas, for example $M=8$, our proposed detector can achieve very satisfactory spoofing detection performance.

In Fig. \ref{fg:Pevspdour} we conduct the simulation to illustrate the performance of our detector as a function of spoofing power $P_E$ ranging from $-20\mathrm{dB}$ to $20\mathrm{dB}$. We set the number of Alice's antennas as $M=4$, the pilot power $P_B=5$, the number of pilot (random) bits as $N_p=N_r=16$, $32$ and $64$, and the probability of false alarm as $P_{fa}=0.01$, $0.001$ and $0.0001$, respectively. From Fig. \ref{fg:Pevspdour}, we can conclude that  more power Eve spends on the pilot spoofing attack leads to higher probability of correct spoofing detection. More importantly, even under very weak spoofing signal, for example $P_E=-5\mathrm{dB}$, the performance of our proposed spoofing detector is still higher than 98\% accuracy with sufficient number of pilot $N_p =64$.

\vspace{-0.0 cm}
\vspace{-0.0 cm}

\vspace{-0.4 cm}

\begin{center}
\begin{figure}[t]
\begin{center}
\includegraphics[width=3.6 in]{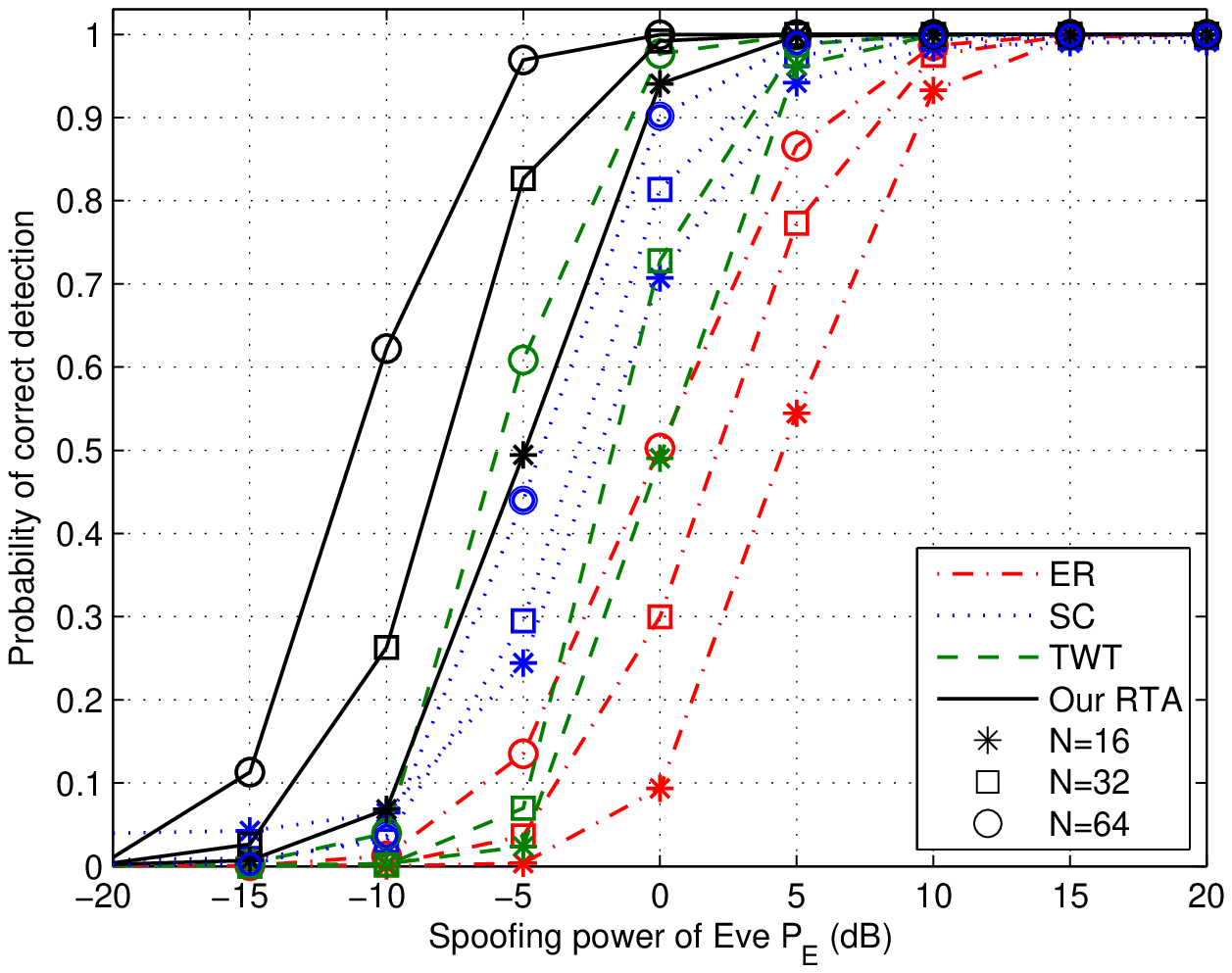} \vspace{-0.5 cm}
\caption{Probability of correct detection versus the spoofing power $P_E$ (training power $P_B=5$, number of Alice's antennas $M=4$, and false alarm rate $P_{fa}=0.001$).}
\label{fg:Pevspdcompare}
%
%
\vspace{0.3 cm}
\includegraphics[width=3.6 in]{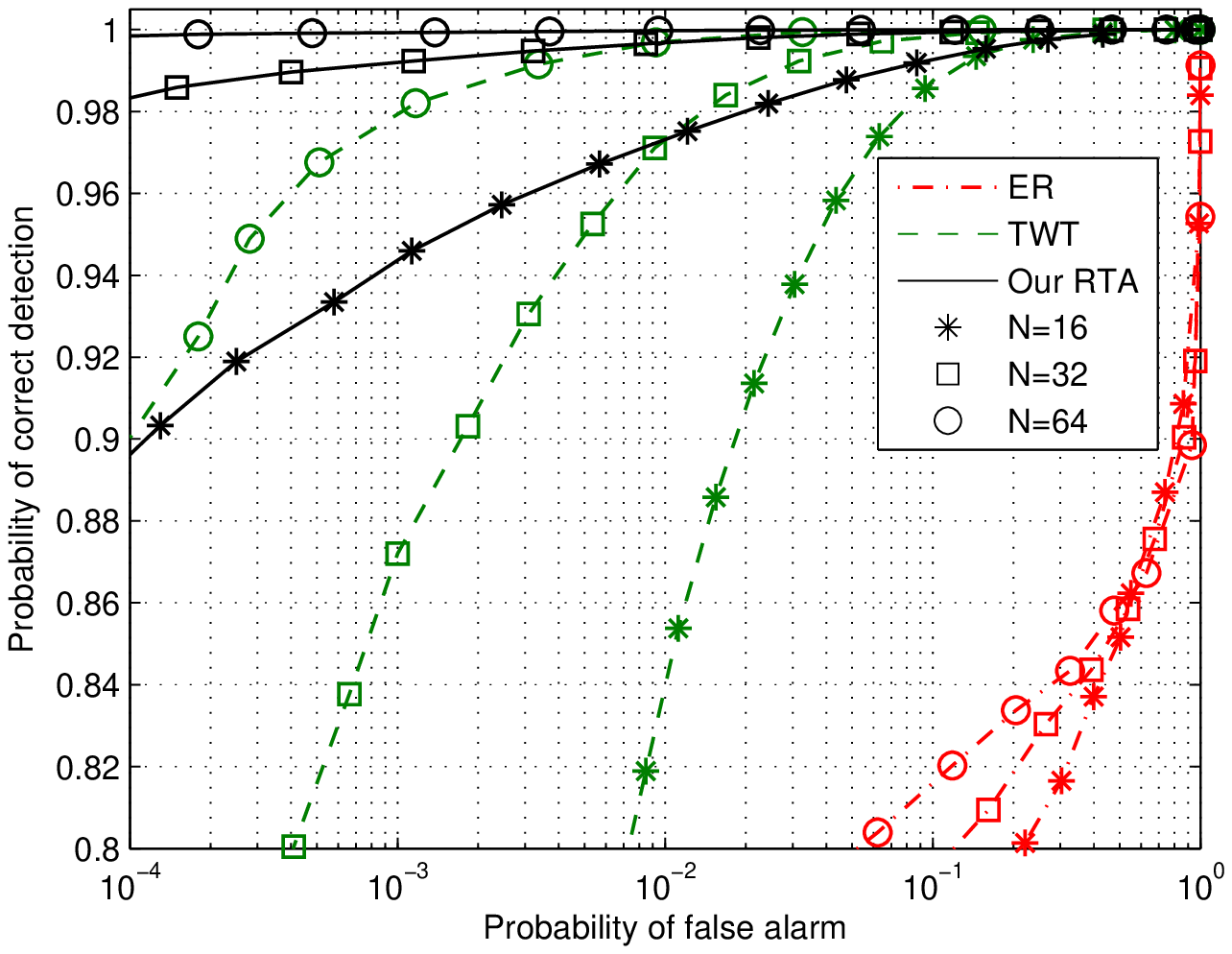} \vspace{-0.5 cm}
\caption{ROC curves of spoofing detectors (training power $P_B=5$, spoofing power $P_E=1$, and number of Alice's antennas $M=4$).}
\label{fg:ROCcompare}\vspace{-0.0 cm}
\end{center}
\end{figure}
\end{center}

In Fig. \ref{fg:Pevspdcompare}, we repeat the similar simulation as Fig.   \ref{fg:Pevspdour} and include several other state-of-the-art detectors for the comparison purposes: \textit{i}) Energy ratio (ER) based approach \cite{ERB}; \textit{ii}) self-contamination (SC) based approach \cite{selfcontamination}; \textit{iii}) two-way training (TWT) based approach \cite{TWTD}. Obviously, our proposed spoofing detector  significantly outperforms its competitors. Next, for the sake of enhanced experimental credibility and comparison results, in Fig. \ref{fg:ROCcompare} we examine the ROC curves of spoofing detectors with the same parameter settings. Again, our proposed spoofing detector is superior to the competitors in all false alarm rate ranges.

\vspace{-0.0 cm}

\vspace{-0.4 cm}

\begin{center}
\begin{figure}[t]
\begin{center}
\includegraphics[width=3.6 in]{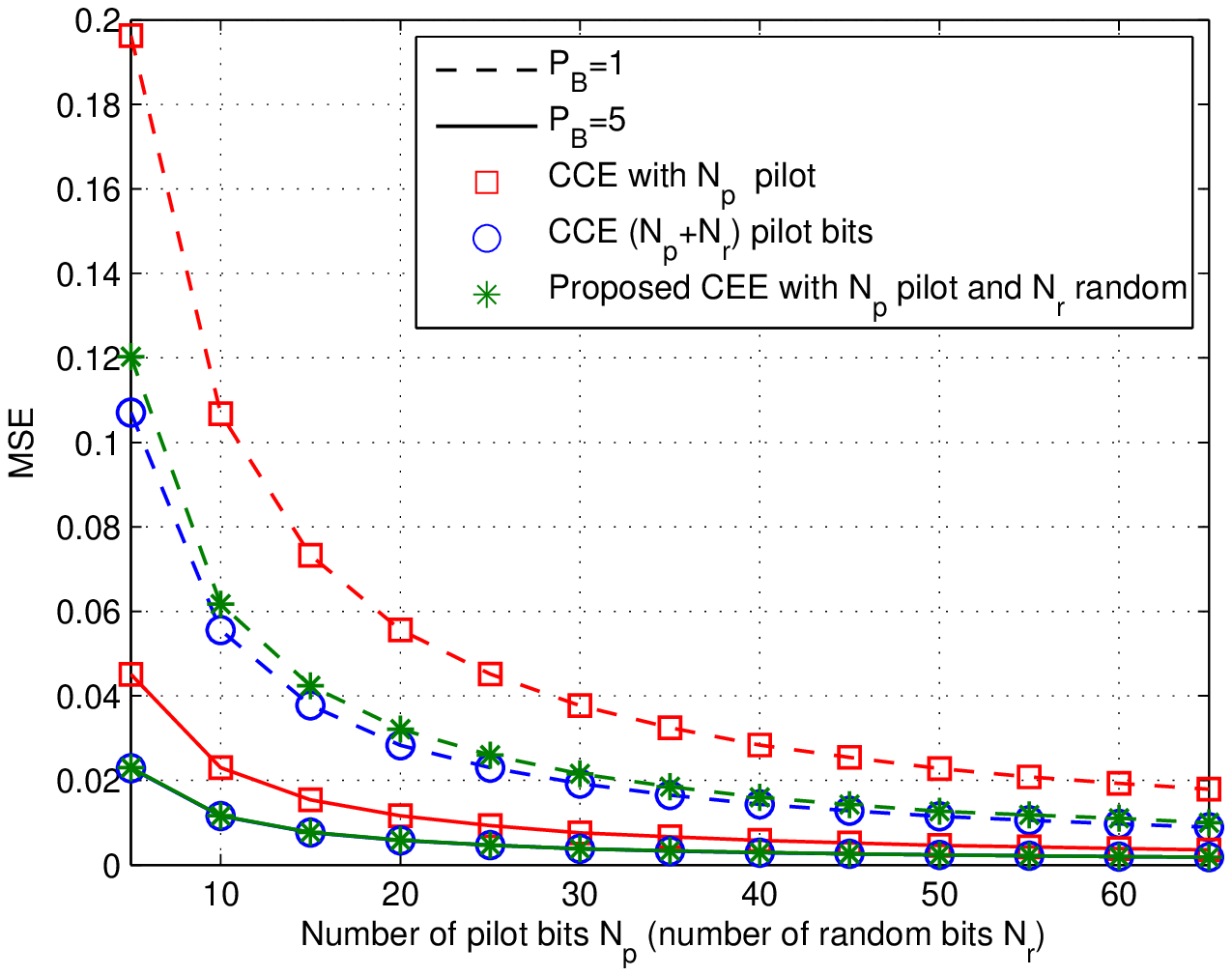} \vspace{-0.5 cm}
\caption{MSE of channel estimation versus the number of the pilot (random) bits $N_p$ (training power $P_B=5$, spoofing power $P_E=1$, and number of Alice's antennas $M=4$).}
\label{fg:MSE_vs_Np}
%
\vspace{0.3 cm}
\includegraphics[width=3.6 in]{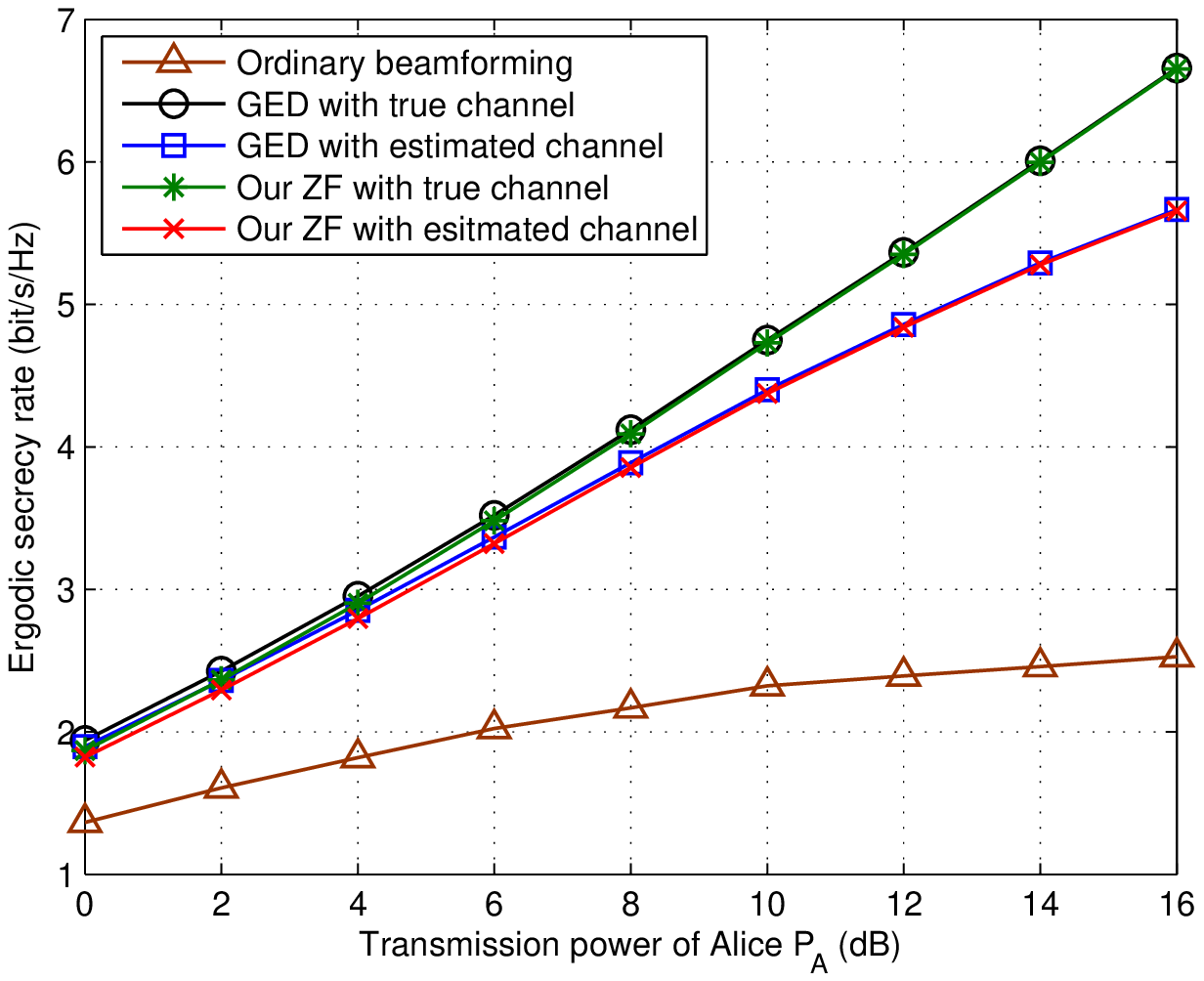} \vspace{-0.5 cm}
\caption{Ergodic secrecy rate versus transmit power $P_A$ (training power $P_B=5$, spoofing power $P_E=1$, and number of Alice's antennas $M=4$, number of pilot (random) bits $N_p=N_r=64$).}
\label{fg:ST_Cs_64} \vspace{-0.0 cm}
\end{center}
\end{figure}
\end{center}

Now we turn to evaluate the performance of the proposed channel estimation enhancement algorithm. In Fig. \ref{fg:MSE_vs_Np} we show mean square error (MSE), which is defined as \be \mathrm{MSE} = \mathbb{E} \left\{ \left\|  {\mathbf{h}_B}/{\| \mathbf{h}_B \|} -  {\widehat{\mathbf{h}}_B}/{ \| \widehat{\mathbf{h}}_B \|} \right\|^2 \right\}, \ee
\nid versus the number of pilot (random) bits $N_p$. Three channel estimation methods are considered: \textit{i}) Conventional channel estimation (CCE) as in (\ref{eq:CCE}) with $N_p$ pilot bits; \textit{ii}) CCE with $(N_p+N_r)$ pilot bits which serves as the performance benchmark; \textit{iii}) our proposed channel estimation enhancement (CEE) approach with $N_p$ pilot bits and $N_r$ random bits. Again, our CEE algorithm uses the same number of pilot bits and random bits $N_p=N_r$. As we can observe from Fig. \ref{fg:MSE_vs_Np}, our proposed CEE can significantly improve the performance comparing with using only $N_p$ pilot bits and approaches the performance benchmark.

Finally, we turn to evaluate the performance of the proposed secure transmission approach. Fig. \ref{fg:ST_Cs_64} shows the ergodic secrecy rate as a function of the transmission power $P_A$. The performances of five transmission schemes are illustrated: \textit{i}) Ordinary beamforming as in (\ref{eq:beamformer}) without any secure effort; \textit{ii}) generalized eigen-decomposition (GED)-based secure  beamforming with true channel which serves as the performance benchmark; \textit{iii}) GED-based secure  beamforming with estimated channel; \textit{iv}) our proposed ZF-based secure beamforming with true channel; \textit{iv}) our proposed ZF-based secure beamforming with estimated channel. We can conclude from Fig. \ref{fg:ST_Cs_64} that our proposed ZF-based secure beamforming can significantly improve the secrecy rate and achieve performance as good as GED-based secure beamforming approach.

\vspace{-0.0 cm}
\vspace{-0.0 cm}

\section{Conclusion}
\label{sec: conclusion}

In this paper, we investigated the problem of detecting pilot spoofing attack for a multi-input single-output (MISO) wiretap system. We proposed a novel random-training-assisted (RTA) spoofing detection scheme which examines the difference of estimated legitimate channels during pilot phase and random phase. Then, for no spoofing case, a simple channel estimation enhancement algorithm was presented to further improve the accuracy of channel estimation. Since the proposed RTA spoofing detection algorithm can obtain the estimates of both legitimate and illegitimate channels,  we also introduced a zero-forcing (ZF)-based secure transmission method when an active eavesdropper is identified. Extensive simulation results illustrated that our proposed RTA scheme can achieve accurate pilot spoofing detection and its performance is superior to other state-of-the-art detectors.

\begin{appendices}
\section{Proof of (\ref{eq:SNR B no spoof}), (\ref{eq:SNR B spoof}) and (\ref{eq:SNR E spoof})}

We first give the proof of (\ref{eq:SNR B spoof}) and then prove (\ref{eq:SNR B no spoof}) and (\ref{eq:SNR E spoof}) according to it. Given $\mathbf{w} =  {\mathbf{\widehat{h}}_B}/{\|\mathbf{\widehat{h}}_B\|}$, we formulate the average SNR at Bob when Eve launches the spoofing attack as
\begin{eqnarray}
\mathrm{\overline{SNR}}_B  \hspace{-0.3 cm} &=& \hspace{-0.3 cm} \mathbb{E}_{\mathbf{h}_B, \mathbf{h}_E} \left\{ \frac{P_A \| \mathbf{h}_B^H \mathbf{w} \|^2}{\sg_B^2} \right\} \non \\
\hspace{-0.3 cm} &=& \hspace{-0.3 cm} \frac{P_A}{\sg_B^2} \mathbb{E}_{\mathbf{h}_B , \mathbf{h}_E} \left\{  \frac{\|\mathbf{h}_B^H \mathbf{\widehat{h}}_B\|^2} {\|\mathbf{\widehat{h}}_B\|^2}  \right\} \non \\
\hspace{-0.3 cm} &=& \hspace{-0.3 cm} \frac{P_A}{\sg_B^2} \cdot
\frac{\mathbb{E}_{\mathbf{h}_B , \mathbf{h}_E } \{ \|\mathbf{h}_B^H \mathbf{\widehat{h}}_B\|^2 \}}   {\mathbb{E}_{\mathbf{h}_B, \mathbf{h}_E}\|\mathbf{\widehat{h}}_B\|^2}.
\label{ap:0}
\end{eqnarray}

We first focus on the derivation of the numerator of (\ref{ap:0}).
Given
\begin{equation}
\mathbf{\widehat{h}}_B = \sqrt{P_B} \mathbf{h}_B + \sqrt{P_E} \mathbf{h}_E + \mathbf{n}_A
\end{equation}
with $\mathbf{n}_A \sim \mathcal{N}(0, \frac{\sg_A^2}{N_p}\mathbf{I}_M)$, we have
\begin{eqnarray}
\mathbb{E}_{\mathbf{h}_B, \mathbf{h}_E} \hspace{-0.1 cm} \{ \|\mathbf{h}_B^H \mathbf{\widehat{h}}_B\|^2 \} \hspace{-0.4 cm} &=& \hspace{-0.4 cm} \mathbb{E}_{\mathbf{h}_B, \mathbf{h}_E} \{ \| \mathbf{h}_B^H(\hspace{-0.1 cm}\sqrt{P_B} \mathbf{h}_B \hspace{-0.1 cm}  + \hspace{-0.2 cm} \sqrt{P_E} \mathbf{h}_E \hspace{-0.1 cm}  + \hspace{-0.1 cm} \mathbf{n}_A)  \|^2 \} \non \\
 &=& \hspace{-0.3 cm}  \mathbb{E}_{\mathbf{h}_B} \{ P_B \mathbf{h}_B^H \mathbf{h}_B \mathbf{h}_B^H \mathbf{h}_B \} +  \non \\
&& \hspace{-0.3 cm}  \mathbb{E}_{\mathbf{h}_B, \mathbf{h}_E} \{ P_E \mathbf{h}_B^H \mathbf{h}_E \mathbf{h}_E^H \mathbf{h}_B \} +  \non \\
&& \hspace{-0.3 cm}  \mathbb{E}_{\mathbf{h}_B} \{ \mathbf{h}_B^H \mathbf{n}_A \mathbf{n}_A^H \mathbf{h}_B \} \label{ap:3}.
\end{eqnarray}
The first term in (\ref{ap:3}) is $P_B\al_B^2M^2$,
the second term in  (\ref{ap:3}) is
\begin{eqnarray}
\mathbb{E}_{\mathbf{h}_B, \mathbf{h}_E} \{ P_E \mathbf{h}_B^H \mathbf{h}_E \mathbf{h}_E^H \mathbf{h}_B \}  \hspace{-0.4 cm} &=& \hspace{-0.4 cm} P_E \mathbb{E}_{\mathbf{h}_B, \mathbf{h}_E} \{ \mathrm{Tr}( \mathbf{h}_B \mathbf{h}_B^H \mathbf{h}_E \mathbf{h}_E^H) \} \non \\
\hspace{-0.4 cm} &=& \hspace{-0.4 cm} P_E \mathrm{Tr}( \mathbb{E}_{\mathbf{h}_B} \{\mathbf{h}_B \mathbf{h}_B^H\} \mathbb{E}_{\mathbf{h}_E} \{ \mathbf{h}_E \mathbf{h}_E^H\} \hspace{-0.0 cm} ) \non \\
\hspace{-0.3 cm} &=& \hspace{-0.3 cm} P_E \mathrm{Tr} (\al_B \mathbf{I}_M \al_E \mathbf{I}_M) \non \\
\hspace{-0.3 cm} &=& \hspace{-0.3 cm} P_E\al_B\al_E M,
\end{eqnarray}
\nid and the third term in (\ref{ap:3}) is
\begin{eqnarray}
\mathbb{E}_{\mathbf{h}_B} \{ \mathbf{h}_B^H \mathbf{n}_A \mathbf{n}_A^H \mathbf{h}_B \} \hspace{-0.3 cm} &=& \hspace{-0.3 cm} \mathbb{E}_{\mathbf{h}_B} \{ \mathrm{Tr}(\mathbf{h}_B \mathbf{h}_B^H \mathbf{n}_A \mathbf{n}_A^H)\} \non \\
\hspace{-0.3 cm} &=& \hspace{-0.3 cm} \mathrm{Tr}(\mathbb{E}_{\mathbf{h}_B} \{ \mathbf{h}_B \mathbf{h}_B^H \} \mathbb{E} \{ \mathbf{n}_A \mathbf{n}_A^H \}) \non \\
\hspace{-0.3 cm} &=& \hspace{-0.3 cm} \mathrm{Tr}(\al_B \mathbf{I}_M \frac{\sg_A^2}{N_p}\mathbf{I}_M) \non \\
\hspace{-0.3 cm} &=& \hspace{-0.3 cm} \frac{M\al_B\sg_A^2}{N_p}.
\end{eqnarray}
Finally, the numerator of (\ref{ap:0}) becomes
\begin{equation}
\mathbb{E}_{\mathbf{h}_B, \mathbf{h}_E} \{ \|\mathbf{h}_B^H \mathbf{\widehat{h}}_B\|^2 \} = P_B\al_B^2M^2 + P_E \al_B \al_E M + \frac{M\al_B\sg_A^2}{N_p}.
\label{ap:numerator}
\end{equation}

Similarly, the  dominator of (\ref{ap:0}) can be derived as
\begin{equation}
\mathbb{E}_{\mathbf{h}_B , \mathbf{h}_E} \|\mathbf{\widehat{h}}_B\|^2 = P_B\al_BM + P_E\al_EM + \frac{M\sg_A^2}{N_p}.
\label{ap:dominator}
\end{equation}

Applying (\ref{ap:numerator}) and (\ref{ap:dominator}) into (\ref{ap:0}), we finally obtain the $\mathrm{\overline{SNR}}_B$ when Eve launches the spoofing attack as below
\begin{equation}
\mathrm{\overline{SNR}}_B = \frac{P_A\al_B}{\sg_B^2} \cdot \frac{MNP_B\al_B + NP_E\al_E + \sg_A^2}{NP_B\al_B + NP_E\al_E + \sg_A^2}.
\label{ap:SNRb spoof}
\end{equation}
\nid Equation  (\ref{eq:SNR B spoof}) is proved.
 $\hfill \square $

In order to prove (\ref{eq:SNR B no spoof}), we let $P_E=0$ in (\ref{ap:SNRb spoof}), then we obtain the $\mathrm{\overline{SNR}}_B$ when Eve does not launch the spoofing attack as below
\begin{equation}
\mathrm{\overline{SNR}}_B = \frac{P_A\al_B}{\sg_B^2} \cdot \frac{MNP_B\al_B + \sg_A^2}{NP_B\al_B + \sg_A^2}.
\end{equation}
\nid Equation  (\ref{eq:SNR B no spoof}) is proved.
$\hfill \square $

Similar to the proof of (\ref{eq:SNR B spoof}), we can obtain the proof of (\ref{eq:SNR E spoof}) by simply changing all the $\al_B$, $\sg_B$, $P_B$, and $P_E$ in (\ref{ap:SNRb spoof}) into $\al_E$, $\sg_E$, $P_E$, and $P_B$, respectively.
Then  the $\mathrm{\overline{SNR}}_E$ when Eve launches the spoofing attack has the expression as below
\begin{equation}
\mathrm{\overline{SNR}}_E = \frac{P_A\al_E}{\sg_E^2} \cdot \frac{MNP_E\al_E + NP_B\al_B + \sg_A^2}{NP_E\al_E + NP_B\al_B + \sg_A^2}.
\end{equation}
\nid Equation  (\ref{eq:SNR E spoof}) is proved.
 $\hfill \square $

\end{appendices}

\end{document}